\documentclass[usenatbib,useAMS]{mn2e}

\usepackage{graphicx,subfigure}

\def\bright{photons s$^{-1}$ m$^{-2}$ arcsec$^{-2}$ $\mu$m$^{-1}$}
\def\lsim{\mathrel{\hbox{\rlap{\hbox{\lower4pt\hbox{$\sim$}}}\hbox{$<$}}}}
\def\gsim{\mathrel{\hbox{\rlap{\hbox{\lower4pt\hbox{$\sim$}}}\hbox{$>$}}}}
\def\lya{Lyman-$\alpha$}

\title[The Case for OH Suppression]{The case for OH suppression at near-infrared wavelengths}

\author[Ellis and Bland-Hawthorn]{S. C. Ellis\thanks{sce@aao.gov.au}$^{1}$ and J. Bland-Hawthorn$^{2}$\\
$^{1}$Anglo-Australian Observatory, P.O. Box 296, Epping, NSW 1710, Australia\\
$^{2}$Institute of Astronomy, School of Physics, University of Sydney, NSW 2006, Australia}

\date{Accepted...... Received .....}

\begin{document}
\maketitle

\begin{abstract}
We calculate the advances in near-infrared astronomy made possible through the use of fibre Bragg gratings to selectively remove hydroxyl emission lines from the night sky spectrum.  Fibre Bragg gratings should remove OH lines at high resolution ($R=10,000$), with high suppression (30dB)  whilst maintaining high throughput ($\approx90$ per cent) between the lines.  Devices currently under construction should remove 150 lines in each of the $J$ and $H$ bands, effectively making the night sky surface brightness $\approx 4$ magnitudes fainter.  This background reduction is greater than the improvement adapative optics makes over natural seeing;  photonic OH suppression is at least as important as adaptive optics for the future of cosmology.

We present a model of the NIR sky spectrum, and show that the interline continuum is very faint ($\approx 80$ \bright\ on the ecliptic plane).  We show that OH suppression by high dispersion, i.e.\ `resolving out' the skylines, cannot obtain the required level of sensitivity to reach the interline continuum due to scattering of light.  The OH lines must be suppressed prior to dispersion.

We have simulated observations employing fibre Bragg gratings of first light objects, high redshift galaxies and cool, low-mass stars.  The simulations are of complete end-to-end systems from object to detector.  The results demonstrate that fibre Bragg grating OH suppression will significantly advance our knowledge in many areas of astrophysics, and in particular will enable rest-frame ultra-violet observations of the Universe at the time of first light and reionisation.
\end{abstract}


\section{Introduction}

\begin{figure}
\centering \includegraphics[scale=0.3,angle=270]{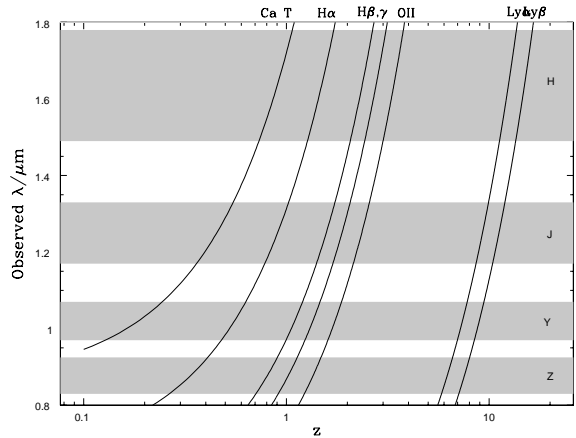}
\caption{The observed wavelength of several useful spectral lines as a function of redshift.  The shaded areas show the wavelength coverage of standard zJH filters.}
\label{fig:observedlambda}
\end{figure}

The future of cosmology is dependent in part on the ability to achieve deep observations at near-infrared  (NIR) wavelengths.  As we strive to observe the early Universe, our observations must be performed at increasingly longer wavelengths, as the diagnostic optical spectroscopic features become more redshifted.  At very high redshifts, observations in the optical become futile due to the redshifted Lyman limit.  Figure~\ref{fig:observedlambda} shows the observed wavelengths of several useful spectroscopic lines as a function of redshift, illustrating the requirement for NIR observations.

The need for NIR observations is well documented.  Over the last few years there have been several major reviews of the current status of our knowledge of the Universe, and in particular of gaps in our understanding, serving as a starting point for strategic plans for future observations (e.g.\ the Australian Decadal Plan for Astronomy 2006-2015\footnote{http://www.atnf.csiro.au/nca/DecadalPlan\_web.pdf}, the UK PPARC roadmap\footnote{http://www.so.stfc.ac.uk/roadmap/rmQuestionsHome.aspx}, the US National Research Council's `Astronomy and Astrophysics in the New Millennium'  \citealt{aanm01} and `Connecting Quarks to the Cosmos' \citealt{cqwc03}). 

These reviews provide similar lists of the `big questions' for the future of astronomy, showing a consensus of opinion at an international level.  Some of the key issues they identify are:
\emph{
What is the nature of dark energy and dark matter?
How and when do the first stars form?
How and when did the Universe reionise?
How and when do galaxies assemble, and how do they evolve?
What is the relationship between super-massive black-holes and galaxies?
How do stars and planetary systems form?
How common are planetary systems capable of supporting life?
}

The success of answering the questions above is reliant on NIR observations.  The first half of the list depends upon the ability to observe the high redshift Universe, whilst the problems concerning planetary systems will also benefit from dust-penetrating NIR observations, especially since planetary sized bodies emit most of their radiation at NIR wavelengths.

Observations of high redshift galaxies carried out in the NIR offer significant advantages over optical observations: extinction due to local dust is smaller in the NIR; $k$-corrections are smaller, and therefore less critical; adaptive optics is easier to achieve in the NIR.  However, there are also significant disadvantages  in NIR astronomy, the chief concern being the high background level.  Table~\ref{tab:skybright} lists typical surface-brightnesses for a good observing site, as used in the ESO ELT exposure time calculator\footnote{http://www.eso.org/observing/etc/bin/ut3/conica/script/eltsimu}.

\begin{table}
\begin{center}
\caption{Typical sky-brightnesses at a good observing site.}
\label{tab:skybright}
\begin{tabular}{ccc}
& \multicolumn{2}{c}{Surface brightness/ mag arcsec$^{-2}$} \\
Band & Natural & No OH emission \\ \hline
U  &     21.5& -- \\
B    &   22.4& -- \\
V      & 21.7& -- \\
R     &  20.8& 21.7 \\
I      & 19.9 &  21.7 \\
J     &  16.0 &  22.1\\
H     &  14.0&  22.2\\
\end{tabular}
\end{center}
\end{table}

At wavelengths longer than the $H$ band, thermal emission from the telescope and surroundings dominate the background emission.  Therefore the detector and camera must be designed to operate at cryogenic temperatures -- a significant technical challenge.  At shorter wavelengths the thermal background is less problematic but instead there is the supplementary frustration of very bright hydroxyl emission (\citealt{mei50}; \citealt{duf51}), resulting from the vibrational decay of an excited OH-radical, formed from the combination of a hydrogen atom with an ozone molecule.  Furthermore the intensity of the OH emission is highly variable (see section~\ref{sec:ohvar} below), making accurate sky subtraction very difficult (e.g.\ \citealt{dav07}).  

The intensity and variability of the NIR background have traditionally restricted the depth of NIR observations compared to optical observations.  However, two attributes of the NIR background spectrum suggest that in principle it should be possible to obtain very deep observations in the NIR.  Firstly the OH spectrum comprises nearly all ($\approx 99.9$ per cent from our model, see \S~\ref{sec:sky}) of the $J$ and $H$ band background.  The interline continuum, due mainly to zodiacal scattered light, is expected to be very faint, e.g.\ the models of \citet{kel98} give a flux density of only $\approx 77$ \bright\ on the ecliptic plane.  Secondly, the OH lines are intrinsically very narrow (\citealt{tur83}), meaning that large sections of the NIR sky spectrum will be very dark if the OH lines can be efficiently removed, thus enabling deep observations.

There are several methods which can, in principle, exploit the dark interline continuum for deep NIR observations, including narrow-band filters which allow observations of a very short bandpass between lines (e.g.\ DAzLE; \citealt{hor04}), or space-based telescopes (e.g.\  JWST; \citealt{gar06}).  Here we discuss ground-based OH suppression, by which we mean the selective removal (by means of low transmission, reflection etc.) of very narrow regions of the spectrum corresponding to the OH lines, whilst maintaining high throughput in the interline regions.  There have been several methods proposed to achieve OH suppression, including spectroscopic masking at high dispersion (e.g.\  \citealt{mai00}), rugate filters (\citealt{off98}) and multi-notch holographic filters (\citealt{blai04}).  We discuss these methods in section~\ref{sec:old}; each offers potential improvements over natural conditions, but in practice there are problems in achieving \emph{efficient} OH suppression by any of these means.  By efficient we mean achieving high suppression of the lines whilst maintaining high interline throughput, for a large number of lines ($\gsim 100$).   In this paper we make the case for using fibre Bragg gratings (FBGs, see \citealt{bland04}) to achieve OH suppression (\S~\ref{sec:fbgs}); FBGs  do not suffer from any of the drawbacks that have prevented successful application of the methods mentioned above.  There is now a real case for believing that highly efficient OH suppression is close to being demonstrated.

Before discussing methods of OH suppression, we first review the background sources contributing to the NIR in section~\ref{sec:sky}.  We then address the scattering properties of dispersive optics (\S~\ref{sec:scatter}), the proper understanding of which is essential to achieve efficient OH suppression.   The study of scattering shows that the OH lines must be suppressed prior to dispersion; the combination of the extremely bright OH lines and the strong scattering wings, that are an unavoidable consequence of any diffraction system, make attempts to remove or `resolve out' the OH skylines after dispersion futile.  We then make the case for OH suppression, i.e.\ selectively removing the OH lines prior to dispersion (\S~\ref{sec:ohsupp}), including a review of previously proposed methods (\S~\ref{sec:old}), and their limitations.   We then show that fibre Bragg gratings offer the prospect of cleanly and efficiently removing OH lines photonically  (\S~\ref{sec:fbgs}).   We calculate the required level of suppression and spectral resolution to achieve scientifically useful observations.  We summarise our case in section~\ref{sec:summary}, and conclude with some remarks on the future in section~\ref{sec:real}.

 \section{The NIR background}
 \label{sec:sky}

We now consider the various elements that contribute to the NIR background, focussing on the wavelength range 0.9--2.0$\mu$m, as the region in which OH suppression will be most effective.  Following Tokunaga in chapter 7 of \citet{allens00}, we divide the NIR background into the following components: the OH airglow; the thermal emission from the ground based telescope;  the atmospheric thermal emission; the zodiacal scattered light; zodiacal emission; galactic dust emission; the cosmic microwave background radiation.  Each of these will be discussed in turn.  Our final model of the NIR background spectrum is shown in Figure~\ref{fig:bgmodel}.

\subsection{Hydroxyl emission lines}
\label{sec:ohlines}

\subsubsection{Emission mechanism}

\begin{figure}
\centering \includegraphics[scale=0.4]{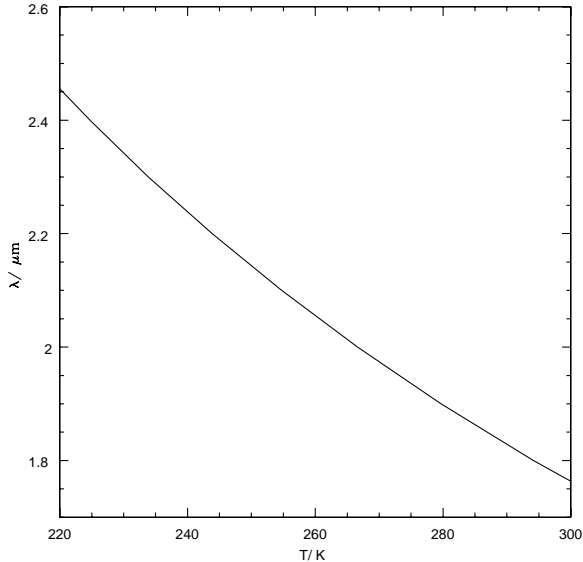}
\caption{A lower limit  on the wavelength at which the OH emission dominates the NIR background as a function of environmental temperature.  This was calculated as the wavelength at which the blackbody spectrum from a typical telescope exceeds that of the zodiacal scattered light at $\beta=60$ degrees.}
\label{fig:tvar}
\end{figure}

The OH emission (\citealt{mei50}; \citealt{duf51}) is by far the largest component of the NIR background at $\lambda \lsim 2.0 \mu$m.  At longer wavelengths the background is dominated by thermal emission, from the telescope and dome etc.  The exact wavelength where the thermal emission becomes dominant depends on the temperature of the environment, as illustrated in Figure~\ref{fig:tvar}, which shows the wavelength at which the blackbody spectrum from the telescope exceeds the blackbody spectrum of the zodiacal scattered light, i.e.\ a lower limit to the wavelength at which OH emission dominates the background.  Thus for very cold sites, such as Dome C in Antarctica with an average year round temperature of $\approx 220K$ (see \citealt{burt05}) OH suppression would be effective at wavelengths $\lambda\lsim 2.4 \mu m$.

The OH emission arises mainly from the combination of hydrogen and ozone from a layer about 9km thick at an altitude of around 87km.  This reaction leaves a rotationally and vibrationally excited OH molecule (and an oxygen molecule) with a radiative lifetime of a few milliseconds (\citealt{con96}, and references therein).  The emission then results from the vibrational decay of the molecule, i.e.,
\begin{equation}
{\rm H + O_{3}  \rightarrow OH^{*} +O_{2}}.
\end{equation}

The intrinsic width of the lines is determined from the  Doppler shift due to the thermal motions of the molecules, from pressure broadening, and from natural broadening.  The former effect dominates the inner width of the line and has a Gaussian form, whereas the latter has a Lorentzian profile, giving the lines broad wings, albeit very faint ones.

The Doppler width can be estimated from the following argument.  For a gas in thermal equilibrium the  gas particles will have a distribution of speeds, but the most probable speed will be when the kinetic energy, $mv^{2}/2$ is equal to the thermal energy $kT$.  Thus the light emitted from the gas particle will be Doppler shifted by an amount $\Delta \lambda/ \lambda = \pm |v_{{\rm r}}|/c$.  Thus,
\begin{equation}
\label{eqn:doppbroad}
\Delta \lambda_{{\rm DB}} = \frac{2\lambda}{c}\sqrt{\frac{2kT}{m}},
\end{equation}
where $m$ is the mass of the particles, and $T$ is the gas temperature.  Thus for OH molecules at a temperature of $T=200$K (\citealt{rous00}), $\Delta \lambda_{\rm DB} \approx 5 \times 10^{-6}\mu$m.

Pressure broadening is the result of interactions or collisions with neighbouring particles in the gas.  The effect can be estimated by considering the average time between collisions or encounters, $\Delta t_{0}$, which is approximately equal to the mean free path divided by the average speed of the particles, i.e.
\begin{equation}
\label{eqn:pressbroad}
\Delta t_{0} \approx \frac{l}{v} = \frac{1}{n\sigma\sqrt{2kT/m}},
\end{equation}
where $\sigma$ is the collisional cross section and $n$ is the number density.  It is assumed that the shape of the line is Lorentzian, which is the result of an emitting atom undergoing simple harmonic motion.    Substituting equation~\ref{eqn:pressbroad} into equation~\ref{eqn:natbroad} yields,
\begin{equation}
\label{eqn:pressbroadwidth}
\Delta \lambda_{{\rm PB}} = \frac{\lambda^{2}n\sigma_{\rm cross}}{c\pi}\sqrt{\frac{2kT}{m}}.
\end{equation}
%
%
We adopt $\Delta_{\lambda_{{\rm PB}}}\approx 3 \times 10^{-7}\mu$m, an upper limit based on the difference in wavenumber  of $<0.002$ cm$^{-1}$ quoted by \citet{tur83}.

Natural broadening is due to the Heisenberg uncertainty in the particles energy,
\begin{equation}
\Delta E = \frac{\hbar}{\Delta t},
\end{equation}
where $\Delta t$ is the duration of the change in energy, such that, 
%
%
%
\begin{equation}
\Delta \lambda_{{\rm NB}}  \approx \frac{\lambda^{2}}{2 \pi c \Delta t}.
\end{equation}
The full calculation shows that the FWHM is in fact twice this value,
\begin{equation}
\label{eqn:natbroad}
\Delta \lambda_{{\rm NB}} = \frac{\lambda^{2}}{\pi c \Delta t}.
\end{equation}
Thus for a vibrationally excited state lasting a few milliseconds, the natural broadening is $\approx 5 \times 10^{-13}\mu$m at $\lambda=1\mu$m.

From the above estimates the Doppler broadening is by far the most dominant contribution to the intrinsic width of the line.  However, due to the much broader wings of the Lorentzian pressure broadening, this mechanism is not negligible, as it will contribute much more to the region between the line peaks.   The natural broadening can be safely neglected.  Therefore in modelling the OH emission line spectrum we include both the Doppler broadening and the Lorentzian profile, i.e. we use a Voigt profile,
\begin{equation}
V(\sigma, \gamma, x)=\int_{-\infty}^{+\infty}\frac{e^{\frac{-x'^{2}}{2\sigma^{2}}}}{\sigma\sqrt{2\pi}} \frac{\gamma}{\pi((x-x')^{2}+\gamma^{2})}{\rm d}x',
\end{equation}
where $\sigma=\frac{\rm \Delta \lambda_{{\rm DB}}}{2\sqrt{2\rm ln}2}$ and $\Delta \lambda_{{\rm DB}}$ is the full-width at half maximum of the Doppler broadened line given by  equation~\ref{eqn:doppbroad}, and $\gamma$ is the half-width at half maximum of the pressure broadened line given by $\frac{\Delta \lambda_{{\rm PB}}}{2}$ in equation~\ref{eqn:pressbroadwidth}.

\begin{figure}
\centering \includegraphics[scale=0.4,angle=0]{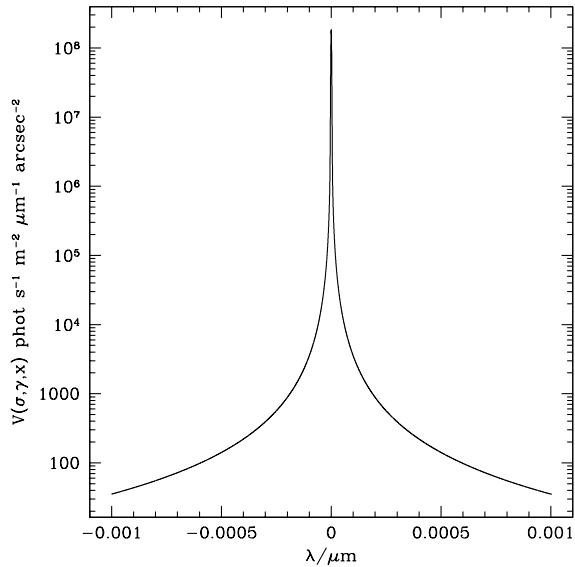}
\caption{The theoretical logged Voigt line profile due to Doppler and pressure broadening, of a very bright OH line line with an integrated flux of 740 $\gamma\ {\rm s}^{-1}\ {\rm m}^{-2}\ {\rm arcsec}^{-2}$ assuming $T=200$K and $\gamma=1.5 \times 10^{-13}$m.  The Lorentzian wings are significant out to at least 1\AA\ from the line centre.}
\label{fig:voigt}
\end{figure}

In our models we assume a temperature at 87km of 200K and $\gamma=1.5\times 10^{-13}$m.  The resulting Voigt profile is shown over a range of $\pm 10^{-3}\mu$m in Figure~\ref{fig:voigt} for the brightest doublets in the $H$ band (which are not resolved).   The plot is logged to emphasise the Lorentzian wings which are significant out to $\approx 5 \times 10^{-4}\mu$m either side of the line centre (c.f.\ zodiacal scattered light described below).  However in practice the interline continuum is dominated by instrumental scatter.  We include this effect in our model, but in practice line widths are dominated by the instrumental profile (see \S~\ref{sec:scatter}) and doublet separations.   

\begin{figure}
\centering \includegraphics[scale=0.4]{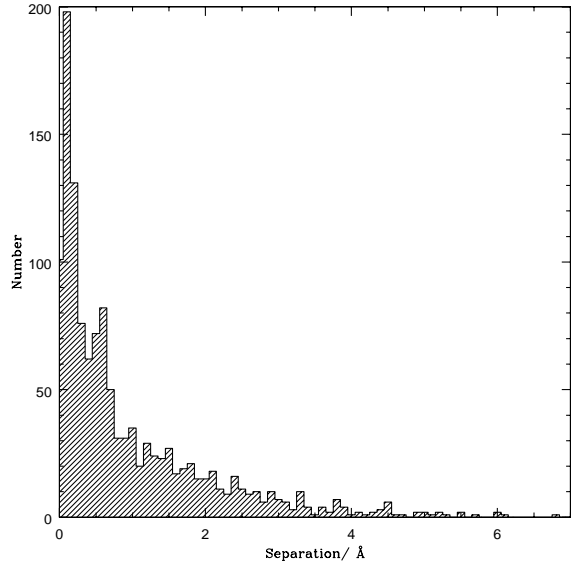}
\caption{The distribution of doublet separations for OH lines with wavelength $0.9$--$1.8\mu$m.}
\label{fig:doublets}
\end{figure}

Note that although the intrinsic width of any individual line is extremely narrow, as shown in Figure~\ref{fig:voigt}, the OH spectrum is actually composed of close doublets.  The doublets are of equal intensity  (\citealt{ost96}) and arise due to $\Lambda$-type doubling (see \citealt{her39}) from the interaction between the rotation of the O and H nuclei and the orbital angular momentum of the electron.  The practical line width is determined by the separation of these doublets.  Using the models of \citet{rous00}, we have determined the doublet separation for all lines in the wavelength range $0.9\le1.8$, shown in Figure~\ref{fig:doublets}.  The distribution is strongly peaked at separations of 0.1--0.2 \AA, which gives a practical limit to the line widths.  Note though, that we model the OH spectrum by treating each line individually.

\subsubsection{Variability}
\label{sec:ohvar}

The OH emission is not only extremely bright, but it is also highly variable in the spatial and temporal domain.  This has significant consequences for the removal of the night sky background from observations.
Despite its strong variability a few trends with time and space are known.  The intensity of the lines increases with zenith distance due to the longer path length through the emitting layer of the atmosphere.  The intensity changes as,

\begin{equation}
\frac{I(\theta)}{I(0)}=\left[1-\left(\frac{ r {\rm sin} \theta}{r+h}\right)^{2}\right]^{-\frac{1}{2}},
\end{equation}

where $\theta$ is the zenith angle, $I(\theta)$ is the intensity at $\theta$, $r$ is the radius of the Earth and $h$ is the altitude of the emitting layer above the ground (\citealt{con96}).

The temporal variation is due to two main causes:  diurnal variation in temperature, and density waves in the upper atmosphere.

\begin{figure}
\centering \includegraphics[scale=0.4]{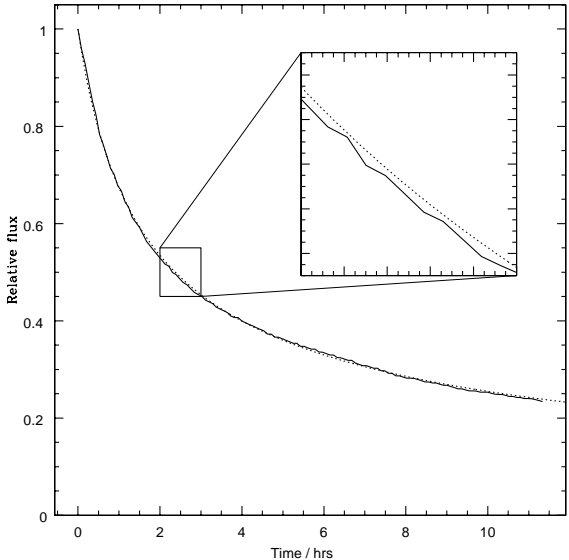}
\caption{The night time variation of the model of \protect\citet{shi70} by the continuous line, and our analytic approximation by the dotted line.}
\label{fig:diurnaloh}
\end{figure}

The diurnal variation has been modelled by \citet{shi70}, taking into account reactions between 14 molecular species and diffusion.  We show their model in Figure~\ref{fig:diurnaloh}, and summarise their description of the variations.  The features result from two emission layers, a broad layer at 45km altitude and a narrow layer at 80km.  The height of the lower layer rises in the afternoon, and partially disappears around sunset, causing a dip in the OH emission.  About twenty minutes after sunset,  both layers rise and the intensity of the emission is greatly enhanced due to an increase in the ozone content at sunset; this causes the sudden rise in the models.  After sunset the lower layer disappears completely because of the removal of hydrogen, and the upper layer decreases slowly due to the decrease of O$_{3}$.  There is also  a rapid dimming and brightening around sunrise.  This model is consistent with observations, which show a decrease by a factor $\sim2$ in the OH emission throughout the night (\citealt{ram92}; \citealt{con96}).  The model is well approximated by the function $1/(1+t)^{0.57}$, where $t$ is the time in hours, as shown in Figure~\ref{fig:diurnaloh}.  We assume that a typical nights observing lasts for 8hrs, so for observations longer than than this we reset the OH brightness every 8hrs.

Aside from this long term decline in OH line strength throughout the night, there is also short term variability caused by density waves (\citealt{ram92}; \citealt{fre00}).  These are typical periodic variations of length 5 -- 15 minutes, with an amplitude of about a tenth of the flux.  

We assume that all OH lines vary in unison for our simulations.  In fact the real situation is more complicated; lines within the same transition will be correlated, but separate transitions will not.   However, to a reasonable approximation, lines from a particular transition all lie within the same region of the spectrum, with very little contribution from other transitions (\citealt{dav07}).  Hence the assumption of unison will be true for sections of spectrum $\sim 0.1\mu$m long.

\subsubsection{Line list}

\begin{figure}
\centering \includegraphics[scale=0.4, angle=0]{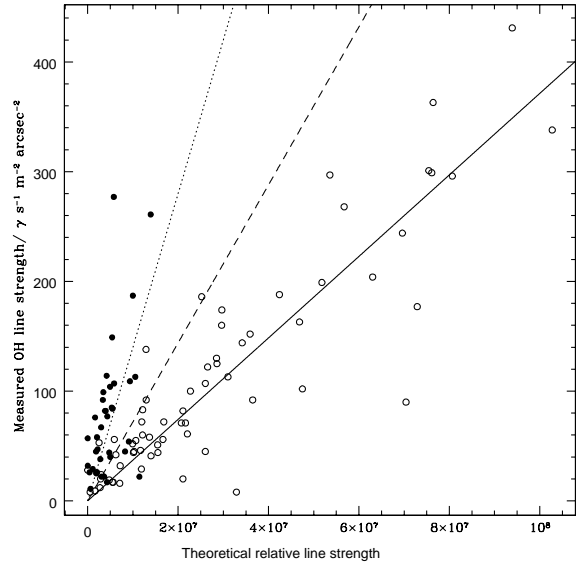} \\
\caption{The correlation between the theoretical relative line strengths of \protect\citet{rous00} and the measured line strengths of \protect\citet{mai93}.  The closed symbols are for the J band and have a slope of $1.4 \times 10^{-5}$.  The open symbols are for the H band, with a slope of $3.7 \times 10^{-6}$.  The dashed line between the J and the H band correlations is the geometric mean of $7.2 \times 10^{-6}$, which was used to calibrate the Rousselot list.}
\label{fig:norm}
\end{figure}

\citet{rous00} have modelled the emission spectrum of the night-sky OH lines, giving accurate wavelengths for 4732 lines, and approximate relative intensities.  In order to use this atlas we have normalised the theoretical relative intensities to the fluxes measured by \citet{mai93} (after first removing the O$_{2}$ lines).  The lines were measured with a resolving power of $R\approx 17,000$ and remained unresolved.  Since the OH lines are close doublets we combine all theoretical lines which have a separation of $\Delta \lambda < \lambda/17000$, before comparing the theoretical and observed line strengths.  Figure~\ref{fig:norm} shows the correlation between the theoretical line strengths and the measured line strengths.   There is a significant difference between the J and H band data, and the H band data shows a stronger correlation.  This is indicative of the transient nature of the OH lines.  We have used the geometric mean of the J and H band correlations to calibrate the Rousselot data.

\subsection{Other lines}

There are a number of O$_{2}$ and N$_{2}$ lines in the J band that are not included in the models of \citet{rous00}.  These lines are self-absorbing and dim rapidly after sunset (\citealt{con96}).  We have included line strengths for these taken directly from \citet{mai93}.  For simplicity, we have assumed they are broadened according to exactly the same prescription as the OH lines.

\subsection{Thermal emission from the telescope}

There will be significant thermal emission from the telescope and the instrument employed.  The instrument design can be optimised to cool as many parts as possible thereby minimising its contribution to the background, but some parts, e.g. optical fibres may be external to the main instrument dewar and therefore unable to be cooled.  Even so, the main contribution will come from the much larger telescope, and we model the emission as a black-body spectrum at $T=273$ K, with an emissivity of $\epsilon=0.02$ (chapter 7, \citealt{allens00}), i.e.,
\begin{equation}
B_{\lambda}=\epsilon\frac{2hc^{2}}{\lambda^{5}(e^{\frac {hc}{k\lambda T}}-1)},
\end{equation}
which can be expressed as a surface brightness,
\begin{eqnarray}
\label{eqn:bb}
N_{\phi}&=&\epsilon\lambda_{\mu {\rm m}}\frac{B_{\lambda}}{hc} \nonumber \\
&\approx&\frac{1.41 \times 10^{16} \epsilon}{\lambda_{\mu {\rm m}}^{4}(e^{\frac{14387.7}{\lambda_{\mu {\rm m}}T}}-1)},
\end{eqnarray}
in units of $ {\rm photons}\ {\rm s}^{-1}\ {\rm m}^{-2}\ \mu{\rm m}^{-1}\ {\rm arcsec}^{-2}$.

\subsection{Zodiacal scattered light}
\label{sec:zodi}

Zodical light is emission from sunlight scattered by interplanetary dust.  As such it has a solar spectrum, multiplied by the scattering efficiency as a function of wavelength.  Individual measurements in the optical (e.g.\ \citealt{lev80}) and the NIR (e.g.\ \citealt{nod92}; \citealt{matsu96}) seem to show a solar spectrum when considered individually, but have different normalisations when compared, with the NIR zodiacal spectrum being twice as bright due to the increased scattering efficiency off larger dust particles at longer wavelengths (\citealt{matsu96}). 

\begin{figure}
\centering \includegraphics[scale=0.4, angle=0]{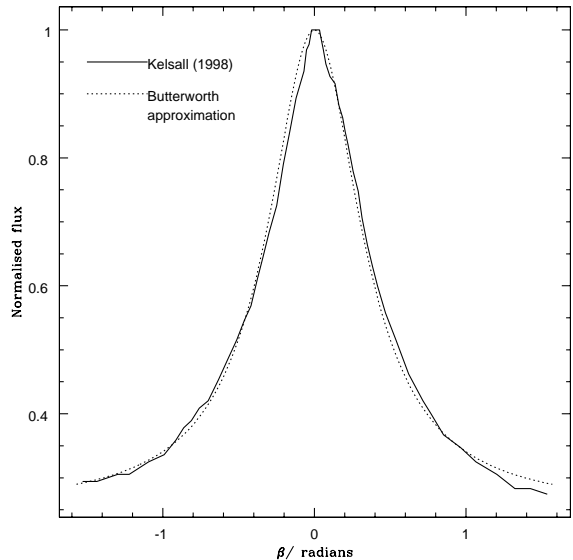}
\caption{The variation of  zodiacal light with ecliptic latitude from \protect\citealt{kel98} shown by the continuous line and our model as given by equation~\ref{eqn:ecl} shown by the dotted line.}
\label{fig:ecliptic}
\end{figure}

The brightness of the zodiacal light obviously decreases as a function of ecliptic latitude.  \citet{kel98} have modelled the zodiacal light in detail in order to remove this component when analysing data from the COBE Diffuse Infrared Background Experiment (DIRBE).  We approximate the Kelsall model with a Lorentzian function, which enables a faster calculation whilst maintaining most of the accuracy.  In Figure~\ref{fig:ecliptic} we show their model's dependence on ecliptic latitude, $\beta$,  for 1.25$\mu$m emission and our fit to it given by,

\begin{equation}
\label{eqn:ecl}
I=\frac{0.75}{(1 + (\frac{2\beta}{0.743})^2)} + 0.25,
\end{equation}

where $\beta$ is in radians.

For the majority of astronomical applications of FBGs we anticipate a preference for high \emph{galactic} latitudes, and therefore a wide range of ecliptic latitudes depending on the galactic longitude, which could change the zodiacal brightness by a factor of $\sim 2$.

For the purposes of the model we normalise a solar spectrum (using equation~\ref{eqn:bb} with $\epsilon=1.08 \times 10^{-13}$ and $T=5800$K) to the \citet{kel98} model value for the ecliptic plane.  
Several measurements and models of the zodiacal scattered light exist in the literature.  These are compiled in Table~\ref{tab:zsl}.  Note that while the models predict the zodiacal scattered light should be fainter at longer wavelengths (since it has the form of a $T=5800$K black-body) the measured data disagree with this.  Note too, that the ground based measurements are rather higher than the models.  In section~\ref{sec:scatter} we suggest that for the ground based measurements scattering of OH light may have a significant contribution to these values.

\begin{table*}
\caption{Measurements and model predictions of the zodiacal scattered light compiled from the literature.  
$^{a}$Includes instrumental background.
$^{b}$Model based on COBE DIRBE measurements.
$^{c}$From a STSI report by M. Stiavelli, see http://www.stsci.edu/hst/nicmos/documents/handbooks/current\_NEW/c04\_imaging8.html\#309385}
\label{tab:zsl}
\small
\begin{tabular}{lccccc}
Source & Measurment or& \multicolumn{2}{c}{Flux/ $\gamma$ s$^{-1}$ m$^{-2}$ arcsec$^{-2}$} & Ecliptic Latitude/ \\
& model&  J & H & degrees  \\ \hline
\citet{mai93} & Measurement & -- & 590 & -- \\
\citet{cub00} & Measurement & 1200 & 2300 & -- \\
\citet{tho07} (NICMOS) & Measurement & 72 & 150 & -45 (HUDF) \\
\citet{tho07} (NICMOS) & Measurement & 74 & 220$^{a}$  & 57 (HDF-N)\\
\citet{tho07} (IRTS NIRS) & Measurement & --- & 110 &--\\
\citet{kel98} & Model$^{b}$ & 43 & -- & 46 (Lockman Hole) \\
\citet{kel98} & Model & 99 & -- & 0 \\
NICMOS handbook$^{c}$ & Model & 130 & 77 & 0 \\
NICMOS handbook & Model & 50 & 31 & 45 \\
\citet{sto98} JWST & Model & -- & 3 & 1AU orbit \\
\end{tabular}
\end{table*}

\subsection{Moonlight}

Moonlight is known to be an insignificant source of background with conventional NIR observations.  This is because the OH skylines dominate over scattered moonlight and also because Rayleigh and Mie scattering are both less efficient at longer wavelengths; Rayleigh scattering has a $\lambda^{-4}$ dependence and Mie scattering has a $\lambda^{-1.3}$ dependence.  However in the regime of OH-suppressed observations these assumptions need to be re-evaluated.

\begin{figure}
\centering
\subfigure[$J$ band]{
\centering \includegraphics[scale=0.7]{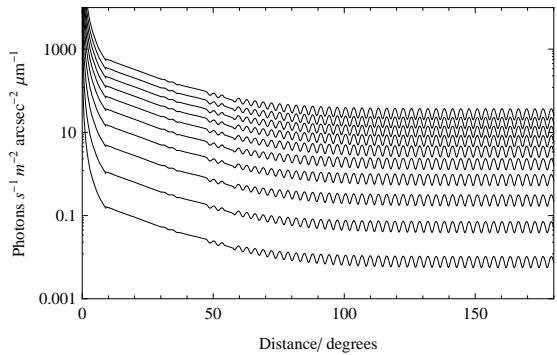}
}
\subfigure[$H$ band]{
\centering \includegraphics[scale=0.7]{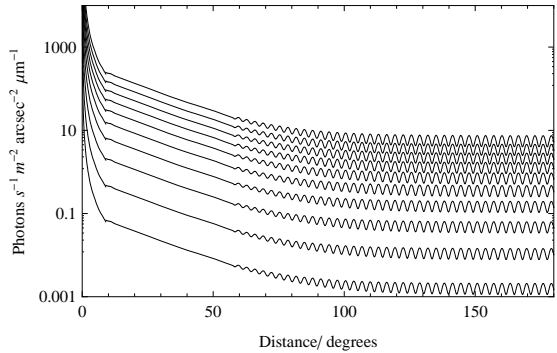}
}
\caption{The brightness of scattered moonlight as a function of angular distance away from the moon.  The different curves are for different lunar phases from 0 to 180 in steps of 20 degrees.  It is assumed that the moon is at the zenith.}
\label{fig:moon}
\end{figure}

\citet{kris91} present a model of the brightness of scattered moonlight that is accurate to 23 per cent.  However, this model is only appropriate for observations in the $V$ band.  Therefore we have adopted the procedure described in the CFHT Redeye Manual\footnote{See http://www.cfht.hawaii.edu/Instruments/Detectors/IR/Redeye/Manual/chapter7.html} to adapt the model to NIR wavelengths.  The principal changes are allowing for the change in colour assuming a solar spectrum, and allowing for the wavelength dependence of the scattering function.  We show the results of the model in Figure~\ref{fig:moon}.  Comparing to estimates of the zodiacal scattered light shows that for observation at a full, or nearly full, moon, then scattered lunar light would be significant out to $\approx 50$ degrees from the moon.  OH suppressed observations should therefore not be performed around the full moon, though `grey nights' should be acceptable.  We note that the effect of scattered moonlight could have had an effect on previous attempts to measure the interline continuum, if these were performed in bright time and the contribution due to moonlight was not taken into account.

\subsection{Other components}

There are a number of other components that make important contributions to the IR background at longer wavelengths ($> 2 \mu$m), but are negligible in the regime considered here.  These are: the thermal contribution from the atmosphere, zodiacal emission from interplanetary dust at the ecliptic pole, Galactic emission from from interstellar dust, the cosmic background radiation.

It is possible that there is also a faint continuum or pseudo-continuum composed of line emission from NO$_{2}$ or other atmospheric models (\citealt{con96}).  Such emission could perhaps explain some of the discrepancy between measurements of the zodiacal scattered light and the ground based measurements of the interline continuum.  However there are no reliable measurements of this continuum, since previous estimates do not take into account the instrumental scattering profile (\S~\ref{sec:scatter}), therefore we do not include any such continuum in our model.

\subsection{Atmospheric transmission}
\label{sec:atmostrans}

\begin{figure*}
\centering \includegraphics[scale=0.8]{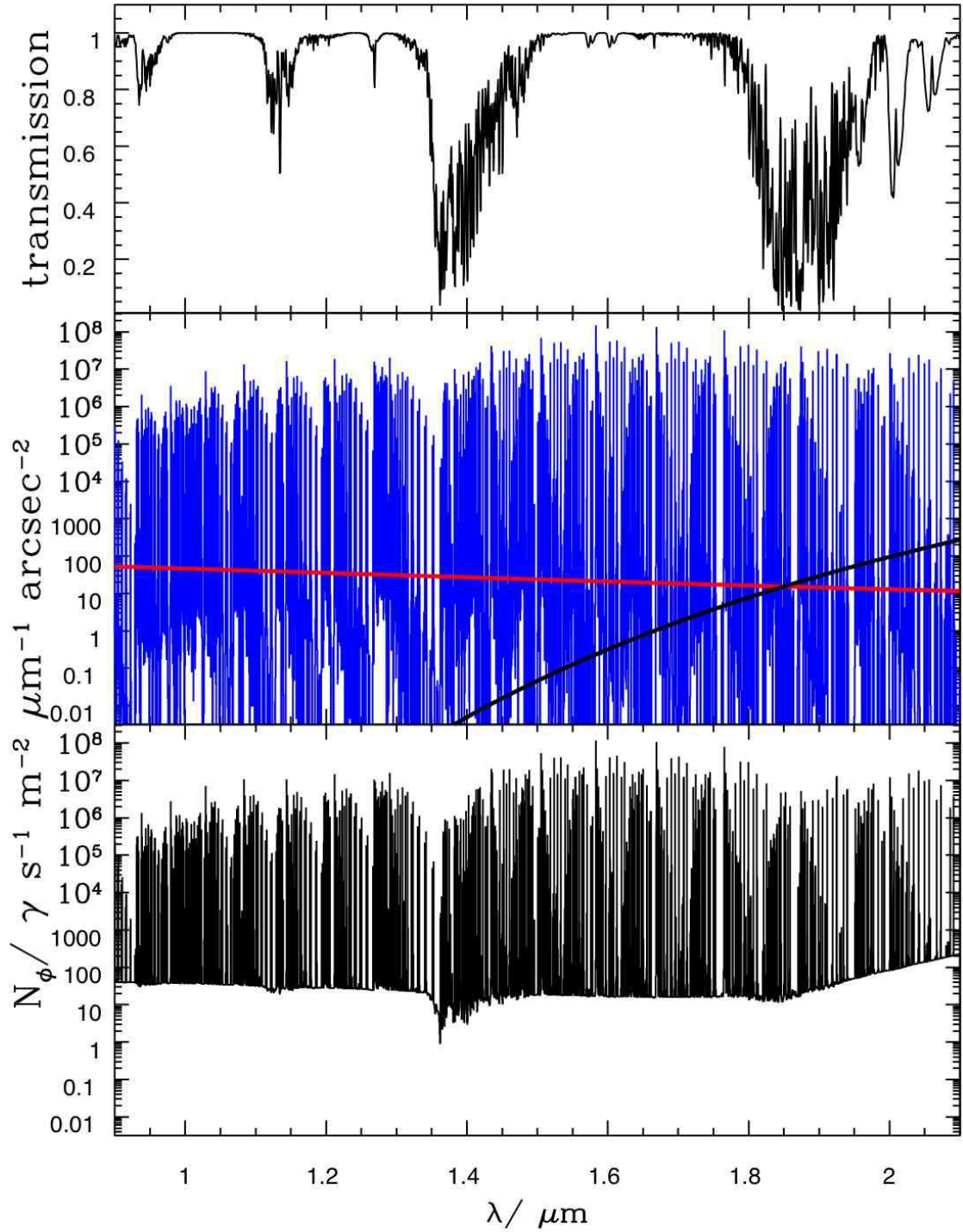}
\caption{The NIR background model.  The top panel shows the atmospheric transmission.  The middle panel show the contributions from the OH spectrum (blue), the zodiacal scattered light (red), the telescope thermal emission (black).  The bottom panel shows the final resulting spectrum.}
\label{fig:bgmodel}
\end{figure*}

The dominant absorption features of NIR light are those due to water, carbon dioxide, nitrous oxide, methane and ozone in the atmosphere, which define the J, H and K band photometric windows.  We have used the model transmission profile at Mauna Kea assuming an airmass of 1.0 and a Water vapour column of 1.6mm in our default sky background model (\citealt{lor92}, obtained from Gemini Observatories\footnote{http://www.gemini.edu/sciops/ObsProcess/obsConstraints/ocTransSpectra.html}).  See Figure~\ref{fig:bgmodel}.

 \section{Scattering of dispersive optics}
 \label{sec:scatter}
 
Because the OH lines are so bright compared to the interline continuum (see Figure~\ref{fig:bgmodel}) they cannot be cleanly removed through high dispersion spectroscopy.  This is because even a small amount of scattered OH light will swamp the true interline continuum.  To see this,  consider the example of a ruled diffraction grating.  The theoretical normalised intensity distribution of a grating with $N$ lines is given by,
\begin{equation}
\label{eqn:grating}
I=I_{0}\left(\frac{{\rm sin}\beta}{\beta}\right)^{2}\left(\frac{{\rm sin}N\gamma}{N{\rm sin} \gamma}\right)^{2},
\end{equation}
where,
\begin{eqnarray}
\beta=\frac{kb}{2}{\rm sin}\theta,\\
\gamma=\frac{kh}{2}{\rm sin}\theta,
\end{eqnarray}
and $k$ is the wavenumber, $b$ is the slit width, $h$ is the slit separation, and $\theta$ is the angle through which the light is diffracted.

\citet{woo94} show that equation~\ref{eqn:grating} may be represented as a Lorentzian plus a constant background,
%
%
%
%
%
%
\begin{equation}
\label{eqn:difflorentz}
Y_{\rm fit}=\frac{\omega^{2}}{(\lambda-\lambda_{0})^{2}+\omega^{2}}+B,
\end{equation}
where $Y_{{\rm fit}}$ is the ratio of intensity at wavelength $\lambda$ to intensity at $\lambda_{0}$, $B$ is a constant background term due to Rayleigh scattering and 
\begin{equation}
\omega=\frac{\lambda_{0}}{N_{\rm e}\pi\sqrt{2}},
\end{equation}
and it is assumed that  $\lambda-\lambda_{0} << \lambda_{0}$ and $N\gsim 100$.   Equation~\ref{eqn:difflorentz} is normalised by a factor 1/($\pi\omega + 2hB$), where $h$ is the groove spacing.

The most important factor in determining the line contribution is $N_{\rm e}$ the number of \emph{effective} lines.  In theory the number of effective lines should be the number of lines illuminated.  However, \citet{woo94} find that $N_{\rm e}$ is always measured to be much lower than this theoretical limit.  In practice $N_{\rm e}$ can be taken to be the smallest number of consecutive grating grooves with no imperfections.  \citet{woo94} report values that are typically 0.25--0.5 the theoretical upper limit, but occasionally as low as 0.16.  Further from the line centre the constant background $A_{\rm B}$ also becomes important.

\begin{figure}
\subfigure[AAOmega, $a=0.97$ and $f=5.5$.]{
\centering \includegraphics[scale=0.4]{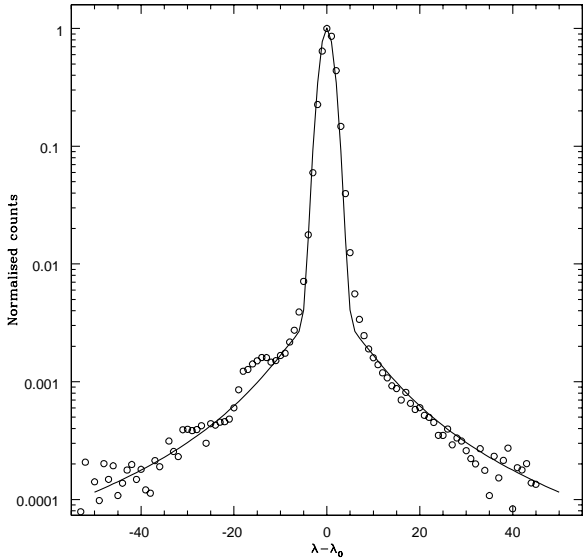}
}
\subfigure[IRIS2, $a=0.89$ and $f=4.0$.]{
\centering \includegraphics[scale=0.4]{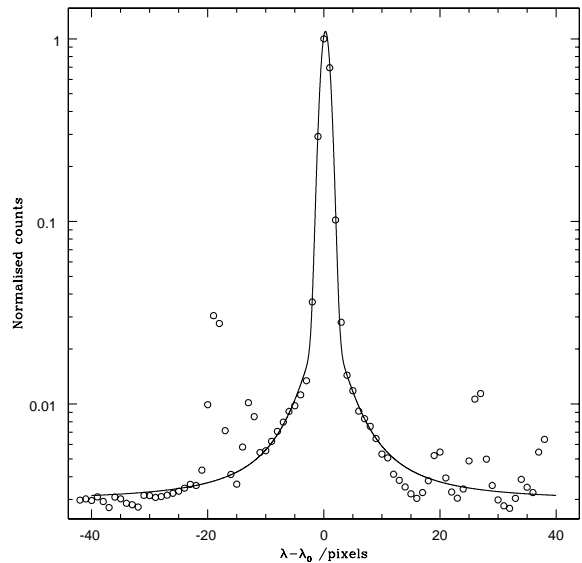}
}
\caption{The line profiles of two spectrographs, AAOmega, a VPH spectrograph and IRIS2 a grism spectrograph.  The measured data are shown by the open circles, and a fit of the form of equation~\ref{eqn:arcfit} is shown by the line.}
\label{fig:aaomegaarc}
\end{figure}

\citet{woo94} fit profiles to 10 different types of gratings including holographically and mechanically ruled gratings and find the scattered light is indeed well fit by a Lorentzian plus 
constant.  Our own measurements with IRIS2 (a grism spectrometer) and AAOmega (a VPH spectrometer) arc spectrum confirm this; we find that the spectral line profile can be fit by a function,
\begin{equation}
\label{eqn:arcfit}
I=I_{0} \left(\frac{e^{-\frac{(\lambda-\lambda_{0})^2}{2 \sigma ^2}} a}{\sqrt{2 \pi } \sigma }+\frac{(1-a) f \sigma  \sqrt{2 \log (2)}}{\pi  \{(\lambda-\lambda_{0})^2+2 f^2 \sigma ^2 \log (2)\}}\right),
\end{equation}
which is simply a Gaussian plus a Lorentzian, with a contribution of factor $a$ from the Gaussian and $1-a$ from the Lorentzian, and the Lorentzian has FWHM $f$ times larger than that of the Gaussian.   We find typical values of $f=4$--6 and $a=0.89$--0.97.    In Figure~\ref{fig:aaomegaarc} we show a fit to the IRS2 and AAOmega arc spectra.

We now show the effects of these scattering wings on the OH spectrum, and in doing so we address the large discrepancy between the ground-based observations of the interline continuum listed in Table~\ref{tab:zsl} and the model values.  We show the effect of scattering on the OH lines around $\lambda=1.665 \mu$m, where \citet{mai93} measured the interline continuum.

We have modelled the scattering from the University of Hawaii 2.2m Coud{\'e} spectrograph, with the 600 lines/mm grating, and $\lambda_{\rm blaze}=1.32\mu$m used for the H band observations by \citet{mai93}.  We have assumed $B=10^{8}$, the value for the Bausch \&\ Lomb grating used by \citet{woo94}.

We use equation~\ref{eqn:difflorentz} to model the scattering profile as it can be normalised exactly.  Note that this will provide a slightly conservative estimate of the scattering far from the line centres, as the wings in equation~\ref{eqn:difflorentz} drop off faster than in the full equation.  

\begin{figure}
\centering \includegraphics[scale=0.3,angle=270]{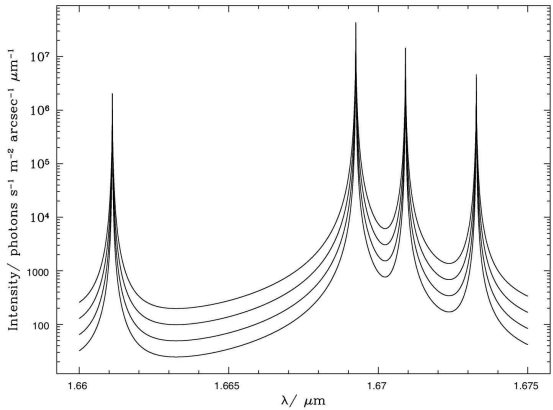}
\caption{The scattered light due to all measured OH lines in the H band.  The different curves are for values of  $N=120,000$, 60,000, 30,000 and 15,000.  The scattered wings are higher for fewer lines.}
\label{fig:scatcont}
\end{figure}

In Figure~\ref{fig:scatcont} we show the scattered light from all H band lines reported in \citet{mai93} in the window 1.660--1.675 (c.f.\ their figure~2) whence the continuum emission was measured.  It is unclear what the theoretical value of $N_{\rm e}$ should be, i.e.\ the number of illuminated lines, but it will be much less than $120,000$ (e.g.\ \citealt{woo94}).  The real value of $N_{\rm e}$, i.e.\ the smallest number of consecutive lines without imperfections will be smaller again.  Therefore we show plots for four different values of $N_{\rm e}$: for $N=120,000=20$cm $\times 600$ lines mm$^{-1}$, i.e. the absolute maximum number of lines, and for values of $N_{\rm e}=60,000$, 30,000 and 15,000.  The scattered wings are higher for fewer lines.  These estimates give a scattered light intensity of between 35 and 280 photons s$^{-1}$ m$^{-2}$ arcsec$^{-2}$ $\mu$m$^{-1}$ at 1.665$\mu$m.  Thus it is possible that scattered light could account for a significant part of the difference between our predicted value of the continuum at $\lambda=1.665\mu$m and that measured by \citet{mai93}, if the number of imperfections in the grating is high.

\begin{figure}
\centering \includegraphics[scale=0.3,angle=270]{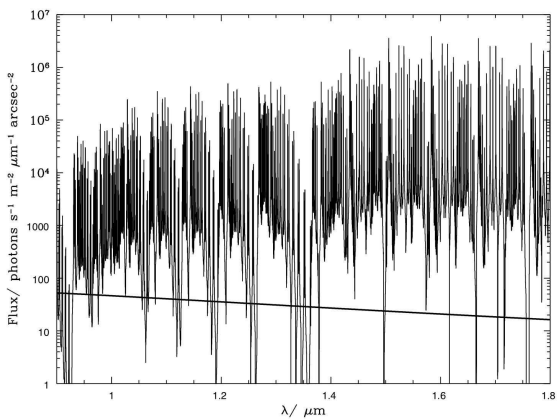}
\caption{The relative strength of the OH spectrum at $R=10000$ (thin line) compared to the zodiacal scattered light at $\beta=60$ degrees (thick line).  The scattered OH line light reaches the interline continuum for 11 per cent of the spectrum.}
\label{fig:skyplot_0_10000}
\end{figure}

The scattering behaviour of dispersive optics thus prevents observations reaching  the level of the interline continuum regardless of instrument background or spectral resolution.  For example in Figure~\ref{fig:skyplot_0_10000} we show the relative strengths of the OH spectrum with a resolution of $R=10000$ (using a scattering profile of the form equation~\ref{eqn:arcfit} with $a=0.90$ and $f=4$) and the interline continuum at an ecliptic latitude $\beta=60$ degrees over the range $0.9\le\lambda\le1.8 \mu$m.  The interline continuum is reached for only 11 per cent of the spectrum between $0.9\le\lambda\le1.8 \mu$m, and 3 per cent between $1.4\le\lambda\le1.8 \mu$m.

 \section{The case for OH suppression}
 \label{sec:ohsupp}
 
By OH suppression we mean the selective removal of OH lines from the incident light whilst allowing light either side of the deleted line to pass through with high efficiency (as opposed to OH avoidance by narrow band filters e.g.\ \citealt{hor04}, which selects light between OH lines, and is therefore restricted to narrow bandpasses).  We discuss how this may be achieved using fibre Bragg gratings below (\S~\ref{sec:fbgs}).  First however, we address the desired performance of a general OH suppression system. 

The OH spectrum contains hundreds of bright lines (\citealt{rous00}; see \S~\ref{sec:sky}).  An OH suppression system must remove a section of finite wavelength, $\Delta \lambda$, from the incident spectrum for each line to be suppressed.  After suppression, these regions will be unsuitable for any scientific analysis.  Therefore OH suppression comes with the penalty of reducing the useful bandpass of the spectrum, which will vary with the number of lines suppressed and the size of the `notches', $\Delta \lambda$.  In practice the regions around bright OH lines are not usually suitable for scientific analysis in any case, since the variation of the OH emission hampers the success of sky-subtraction and residual sky lines are usually observed.

In designing an OH suppression system it is thus necessary to choose how many lines to remove and at what resolution in order to maximise the efficiency of the system.  We now address this issue for notches fixed at $\Delta \lambda=$1\AA , 3\AA\ and 10\AA\ ($\lambda/\Delta \lambda \approx 10000$, 3000 and 1000), for varying numbers of lines.  The lines are suppressed in order of their relative brightness, with the brightest lines removed first.  We measure the efficiency of the system by the fraction of useable spectrum remaining and the brightness of the faintest line suppressed (as an estimate of the expected sensitivity of  the observations).  Note that in practice many OH lines are very close doublets, and therefore these can be removed with a single 1\AA\ notch.

\begin{figure*}
\begin{minipage}[c]{0.5\textwidth}
\centering \includegraphics[scale=0.43,angle=0]{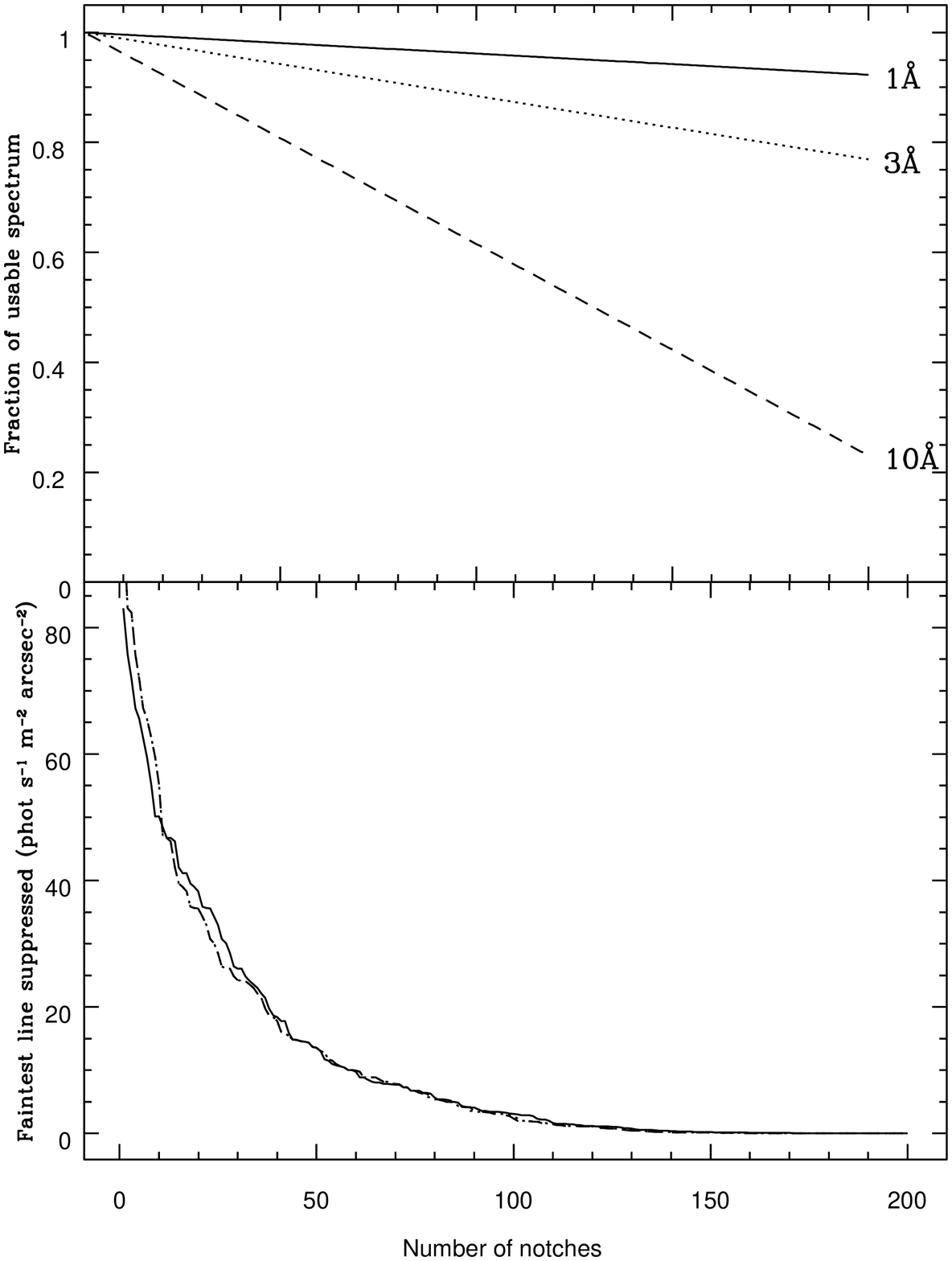}
\end{minipage}%
\begin{minipage}[c]{0.5\textwidth}
\centering \includegraphics[scale=0.43,angle=0]{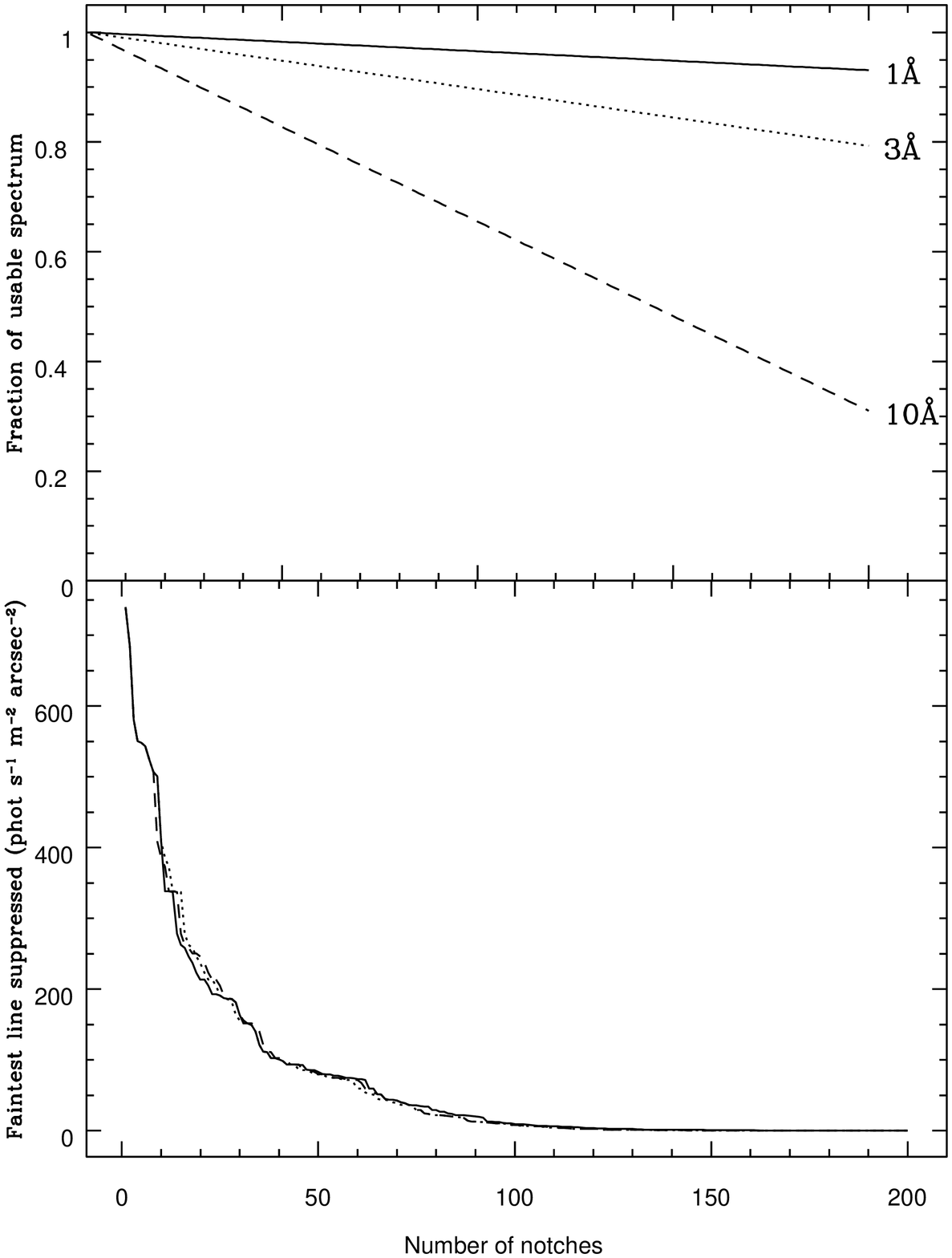}
\end{minipage}
\caption{The effect of increasing the number of OH suppressing notches on the incident spectrum.  The left panels show results from the $J$ band and the right the $H$ band.  The top panels show the fraction of the total passband that remains usable after suppression.  The bottom panels show the flux of the faintest line suppressed.}
\label{fig:nnotches}
\end{figure*}

Figure~\ref{fig:nnotches} shows the results for up to 200 notches across the $J$ and $H$ bands.  Even with 200 lines of width 1\AA, suppressing down to a level of  $4 \times 10^{-3}$ \bright\ (far fainter than would practically be necessary), only 8 per cent of the  $J$ band has been removed, and only 7 per cent for 200 lines in the $H$ band.  Thus, by virtue of the intrinsic slenderness of the OH-emission lines, suppression at high resolution is a highly desirable objective, maintaining a large fraction of scientifically useful wavelength range, whilst vastly increasing signal to noise.

\section{Methods of OH suppression}

There are several methods which have been proposed, and in some cases employed, to achieve OH suppression.  However, to date no instruments exist which can suppress OH light at high efficiency whilst maintaining good interline throughput.  In the next section we will review proposed methods of OH suppression and address their limitations.  This will be followed by a discussion of FBGs, in which we will argue that there is now a real case for believing that highly efficient OH suppression will finally be acheived.

\subsection{Previous methods of OH suppression}
\label{sec:old}

\subsubsection{High dispersion masking}

\citet{mai93b} present a method of OH suppression using high dispersion and a ruled mirror.  The basic idea is to disperse the light at high resolution, whereupon it is reflected from a mirror-mask, which has a series of black lines coincident with the position of the OH lines, which are therefore not reflected.  The reflected light, with the OH lines removed, then proceeds back through the camera and grating system in the opposite direction to its initial journey, and hence is recombined into a `white-light' beam.  This final beam may then be dispersed at low resolution as required.  

This method of mechanically removing the OH light and recombining is preferable to using software to achieve the same result, e.g.\ by removing the OH lines, then rebinning the spectrum, because the read-noise becomes much more significant for a highly dispersed spectrum (\citealt{iwa94}).

Spectrographs employing this technique have been built and used on the Subaru telescope, e.g. OHS+CISCO (\citealt{iwa01}; \citealt{mot02}), and on the University of Hawaii 2.2m (\citealt{iwa94}).  These instruments succeeded in reducing the cores of the OH lines to $\approx 4$ per cent of their original strength.  However, it is the reduction in the background between the lines that is important in assessing the merit of an OH suppression system, since this is where scientific measurements will be made.   Thus the fundamental limit to the performance of OHS-style suppressors is the scattering of the OH light.  This process was described in detail in section~\ref{sec:scatter}.  Figure~\ref{fig:scatcont} illustrates the problem: the Lorentzian scattering wings prevent the interline continuum ever being reached.  This is true even at high resolution as shown in Figure~\ref{fig:skyplot_0_10000}.  Furthermore, in a mirror-mask system the problem of scattered light is compounded, since scattering will occur for each pass through the grating system, and from diffraction from the mirror-mask itself.   The scattered wings of the OH line remain in the spectrum since only the core of the line is suppressed with the mirror mask; between the lines the gains will be much less.
To realise efficient OH suppression this scattering problem must be avoided; the OH  light must be removed prior to dispersion.


The same principle of a mirror-mask design is being used in FMOS (fibre multi-object spectrograph; \citealt{mai00b}), which should begin use on the Subaru telscope in 2008.  The performance of the OH suppressing part of the FMOS instrument is expected to be similar to OHS, but with the advantage of multiplexed observations and improved spectrographs.

\subsubsection{Rugate filters}

Rugate filters have been proposed by \citet{off98} as a method of OH suppression applicable to imaging.   Rugate filters consist of a transparent substrate onto which is deposited a thin film of dielectric material.  By varying the composition of the film as it is deposited the refractive index can be made to vary throughout the coating.  By controlling the refractive index variation any arbitrary transmittance profile may be designed, e.g.\ by combining various sinusoids using Fourier theory.

The chief obstacle in realising OH suppression with rugate filters has been the difficulty of manufacture.  Deposition of each layer of the film requires very tight tolerances in the thickness of the layer, which must be uniform over the entire area of the substrate.  These difficulties have so far made manufacture of satisfactory filters impracticable. 
A further problem is that even should the tolerances of each layer be met, thousands of layers are required which leads to a coating which is tens of microns thick.  The stresses in each layer of so many coatings ultimately makes each layer unstable (\citealt{enn66}).

\subsubsection{Holographic filters}

\citet{blai04} suggest the use of volume phase holographic filters (\citealt{bar00}) as a method of OH suppression.  These filters use holographic recording techniques to record a grating consisting of fringes of varying refractive index in a volume of transparent material, typically dichromated gelatin.  If the fringes are arranged to be parallel to the surface of the filter, no dispersion occurs, and the grating acts as a notch filter.  By controlling the fringe pattern, several notch filters may be multiplexed within a single filter.
\citet{blai04} demonstrate a filter capable of suppressing 10 lines (or close doublets) at 10dB, with a notch width of 1\AA\ and an interline throughput of 85 per cent.

These filters have the advantage as a viable method of OH suppression in that they are currently available for purchase from Photon Etc.  However, there are some difficulties in their use in a practical situation.  The wavelength response of the filter profile shifts as a function of off-axis angle, which restricts the use of the filters to slow beams.  In order to use the filters in slow beams, the notch width must be made larger to accommodate the shifting filter profile, reducing their efficiency.  Another issue is that a single filter is made up from several glass substrates which are cemented together such that the layers between each substrate aligns with the optical axis.  To avoid scattering between the layers the boundaries must be blackened, which therefore divides the beam before it enters the VPH.

A comparison of VPH filters with FBGs strongly favours the latter. The relatively small number of lines per filter compared with FBGs, coupled with the lower interline throughput is disadvantageous.  FBGs are able to suppress 150 lines across a 400nm window with a interline throughput of 88 per cent.  The same 150 lines would require 15 of the filters demonstrated by \citet{blai04}, or 5 of the 30 line filters advertised by Photon Etc.  This would result in a throughput of 9--44 per cent. 

FBGs can suppress the OH lines at 30dB, whereas the VPH filters suppress filters at 10dB.  This makes a significant difference in the performance of the two devices.  In the next section we show how the suppression factor strongly influences the performance of FBGs (Figure~\ref{fig:rsupp}).  If the OH lines are not suppressed strongly enough, the residual lines will still dominate the background signal due to the scattering effects described above.

\subsection{Fibre Bragg gratings}
\label{sec:fbgs}

\citet{bland04} present a radical new approach to OH suppression employing fibre Bragg gratings.  Fibre Bragg gratings (FBGs) are optical fibres with a slowly varying refractive index in the core.  As light propagates down the fibre it suffers Fresnel reflections due to the varying refractive index.  By properly controlling the index variation it is possible to produce strong reflections at specific wavelengths.  \citet{bland04}  showed it was possible
to produce a low loss filter
defined by a series of irregularly spaced, narrow notches with
suppression depths as high as 30 dB, whist maintaining high throughput outside of the notches (losses$\approx 0.03$ per 100nm wavelength coverage). Their design corrects for
18 OH lines in the H band.
 
Two steps are required in order for this technology to be
useful for astronomical observations: the ability to suppress many lines, and the ability to work with multi-mode fibres.  These two requirements are not easily compatible.  When one considers the huge complexity of suppressing up
to 150 OH lines within the H band, the amplitude and phase
structure of the grating demands that this is printed into a
single-mode fibre so that the $\sim 10^{5}$ cycles are
available to produce the filter response along one propagation
axis (\citealt{bland08}). In a multimode fibre, a much
larger number of cycles is required to
achieve the same grating response to each mode, thereby
greatly increasing the complexity of the FBG. 
Thus FBGs achieve 
peak performance within a single-mode fibre (10 micron core),
whereas astronomical fibres, with their larger core diameters,
are highly multi-moded.  Therefore a successful implementation of photonic OH suppression requires
that the many modes of the input fibre be transformed into
the same number of single modes acting in parallel.

To this end, we developed the
photonic lantern (\citealt{leo05}, \citealt{bland05b})
which decomposes the spatial modes of the input fibre into a
parallel degenerate array of single-mode fibres. Identical
gratings are printed into each of the fibres. Once the filtered
light exits the fibre array, these are recombined by an identical
taper back to a multimode output fibre.    \citet{hor07} have demonstrated that fibres with as few as 5 modes can have high coupling efficiency, paving the way for AO fed FBGs in the near future.

The combination of these two innovative technologies means that the development of FBGs for astronomical purposes is now at a stage that is compatible with astronomical requirements.  FBGs are now being developed to remove $\sim 150$ lines in each of the $J$ and $H$ bands.  The brightest lines are suppressed at 30dB, and fainter lines are suppressed by an amount proportionally less.    Photonic lanterns are being developed with 37 fibre tapers.  The resulting multimode FBG will be a complex device, and
expensive to manufacture, at least in the current incarnation.
Initially, only tens to hundreds of fibres are envisaged, but
ultimately, the technology will undergo further development
to bring down the unit cost.

In light of the discussion in section~\ref{sec:scatter}, an important advantage of FBGs is that they can be incorporated into a spectrograph before the dispersive elements, thus the OH light is strongly rejected with very little scattering, offering for the first time the capability to achieve extremely deep NIR observations without the requirement to put a telescope and instrument into space.
 
\begin{figure}
\centering \includegraphics[scale=0.3,angle=270]{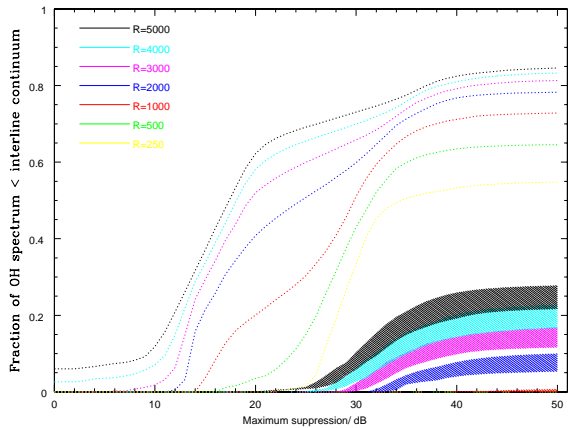}
\caption{The fraction of the $1.4$--$1.8 \mu$m window over which observations can reach the interline continuum as a function of maximum suppression.  Spectral resolutions of $R=250$, 500, 1000, 2000, 3000, 4000 and 5000 are shown by the different colours.  The hatched curves show the range for the \protect\citet{kel98} and NICMOS models at an ecliptic latitude of $\beta=60$ degrees.  The dotted lines are for the interline continuum of \protect\citet{mai93}.}
\label{fig:rsupp}
\end{figure}

The ultimate goal of OH suppression is to reach the interline continuum over a significant fraction of the $J$ and $H$ bands.  The level of suppression required to achieve this depends on the spectral resolving power of the instrument.  Figure~\ref{fig:rsupp} shows the fraction of the $1.4$--$1.8 \mu$m window in which the OH spectrum falls below the zodiacal scattered light as a function of suppression and spectral resolution.  A maximum suppression of 40dB yields significant gains, and thereafter the improvements flatten off.  This is true of both the high interline continuum of \citet{mai93} and the models.

\subsection{Simulations}

In order to assess the expected performance of FBGs we have thoroughly modelled an entire observing system from astronomical source to the detector, incorporating an instrument employing FBGs.    Three different source types were chosen to illustrate the power of OH suppression with FBGs, a $z=11$ QSO, a $z=3$ galaxy and a faint T dwarf.

\subsubsection{System model}
\label{sec:sys}

\begin{figure*}
\centering \includegraphics[,angle=0,scale=0.55]{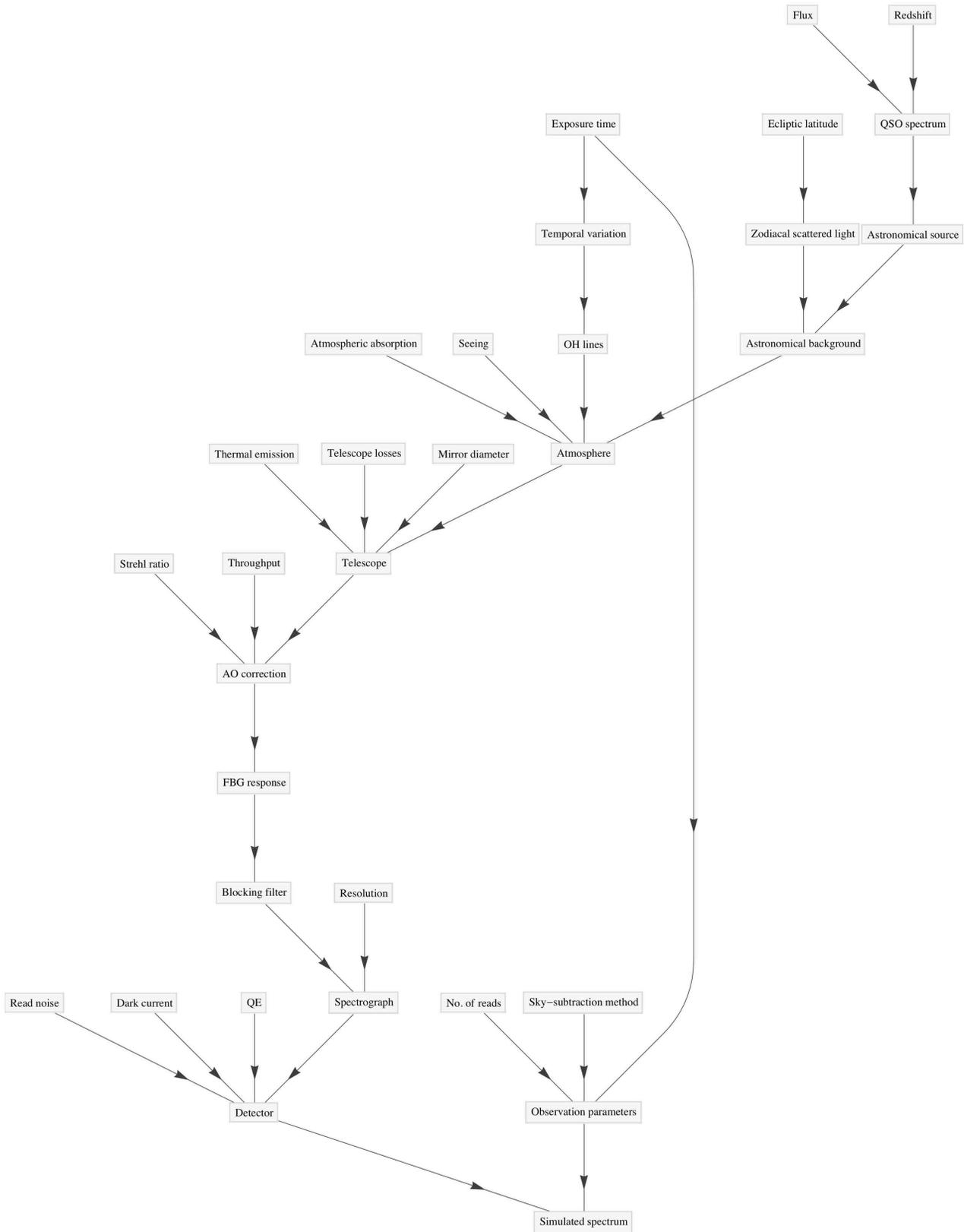}
\caption{A much simplified schematic overview of a typical simulated spectroscopic observation, showing the main causes of emission and losses, with which we have modelled the performance of the fibres. }
\label{fig:schematic}
\end{figure*}

The simulations work by starting with a fluxed 1d object spectrum and modifying it according to background contributions and losses  for each component of the system.  A simplified schematic overview of a typical simulation is shown in Figure~\ref{fig:schematic}, showing the sources of emission and losses we have modelled.  Many of the components shown are optional, for example, we can also model galaxy, and stellar spectra, amongst other sources.  It should also be noted that most of the components shown are fully adjustable, e.g. OH line strengths, zodiacal scattered light etc. can all be adjusted according to meaningful parameters.  The observation parameters and mode of observation are then applied to generate the final spectrum.  

The details of the simulations were as follows.  The zodiacal background is described above (\S~\ref{sec:zodi}), and we modelled observations at $\beta=60$ degrees.  The OH line spectrum was used as described in section~\ref{sec:ohlines}, including temporal variation.  The atmospheric absorption is described in section~\ref{sec:atmostrans}.  We assumed an 8m telescope with obstruction losses of 5 per cent was employed in Cassegrain focus with a reflectivity of 97 per cent off both mirror surfaces.  We modelled the thermal emission from the telescope as a $T=273$K blackbody with an emissivity of 3 per cent.  

The FBGs are modelled as 150 line gratings of 1\AA\ width.  The fibres are taken to have a core size of 25$\mu$m, i.e.\  few mode fibres,  corresponding to an angular diameter of 0.2 arcseconds at f/5.  The fibre coupling losses are $\approx 0.15$ for a few mode fibre of this size (\citealt{hor07}).  These are fed with a generic AO system, achieving a Strehl ratio of 0.3, with a throughput of 0.9.  The fibre aperture losses are thus, $\approx 46$ per cent, due to lost light, assuming that the PSF can be modelled as a diffraction limited PSF plus a Gaussian seeing disc with FWHM 0.5 arcsec, where the two components are normalised such that the diffraction limited part contains a power equal to the Strehl ratio.  

The spectrograph is assumed to have a throughput of 35 per cent, and an adjustable resolution.  The sampling is assumed to be 3 spatial by 3 spectral pixels.  A systematic wavelength calibration error of a tenth of a pixel FWHM has been included.  We assume a blocking filter is used at the entrance to the spectrograph and apply the corresponding transmission.  The detector is taken to have an effective read noise of 4e$^{-}$ when up-the-ramp sampling is employed, and a dark current of two thousandths of an electron per  pixel per second,  typical of Rockwell Hawaii-2RG HgCdTe arrays (\citealt{smi06}).  Observations are  composed of 30 minute reads, with sky-subtraction performed by cross-beam switching between pairs of fibres.

\subsubsection{General results}

The system described above results in a reduction of the total background by a factor of of 42 and 17 in the $H$ and $J$ bands respectively, corresponding to 4 and 3 magnitudes.  \emph{It is important to realise that this reduction applies across the entire spectrum including the interline regions}.  For example at $\lambda=1.665\mu$m (where \citealt{mai93} measured the interline continuum) the background would be dominated by scattered OH light in the absence of FBGs.  However, because FBGs suppress the OH lines before scattering, the interline background is also reduced by a factor of 40.  We emphasise that it is the magnitude of the interline reduction that is scientifically important, and that the advantage of FBGs over other methods of OH suppression is their ability to suppress the OH light before the light enters the spectrograph, and therefore before any significant scattering surfaces are encountered.  Methods to remove OH light after dispersion may achieve a good overall reduction in the overall background due to the removal of the cores of the bright emission lines (as could also be achieved through software suppression), but between the lines, where the science will be done, the reduction will not be as good due to the scattered light.

\begin{figure}
\centering \includegraphics[scale=0.4]{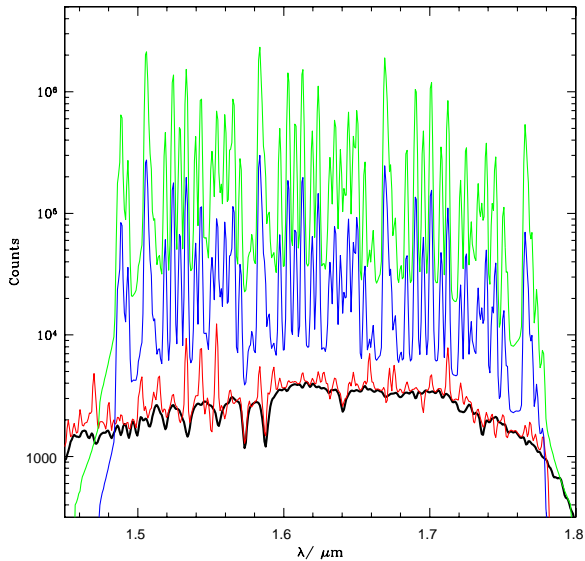}
\caption{Simulated spectra of a $z=3$ $H=19$ mag galaxy, observed at a resolution of $R=1000$ for 1 hr.   These spectra are of a very bright galaxy, and have no sky-subtraction in order to emphasise the equivalent width of the OH lines compared to the continuum.  The thick black line shows the actual object spectrum; the red line shows the object plus background for the galaxy observed with AO and FBGs as described in section~\ref{sec:sys}; the blue line is for AO correction only; the green line is for natural seeing of 0.5 arcsec.}
\label{fig:ao}
\end{figure}

We compare the reduction in background resulting from OH suppression to that resulting from AO correction in Figure~\ref{fig:ao}.    Changing from a 0.5 arcsec aperture for natural seeing to a 0.2 arcsec aperture for AO correction results in a reduction of the background by a factor 6.25.  Applying OH suppression then results in a \emph{further} reduction of a factor of 42 in $H$ or 17 in $J$ as described above.  Given that diffraction limited observations would severely over-sample a high redshift galaxy of size 1 arcsecond, we argue that the main advantage from AO for cosmology is the reduction in background light.  Consequently we argue that OH suppression is at least as important for cosmology as AO, since it produces a greater level of background suppression.  Of course both technologies must proceed together, but given the significant advances promised by OH suppression we argue that the development of FBGs should be given a very high priority for future instruments on existing telescopes and on the ELTs of the future.

\subsubsection{QSO simulation}
\label{sec:qsosim}

The typical surface brightness of sources responsible for reionisation is AB=27--29 mag arcsec$^{2}$, achievable by  a source of 26 AB mag per square arcminute if the number density is $\approx 0.1$--10  arcmin$^{-2}$ (\citealt{sti04}).  This source density is reasonable considering the number counts of \citet{yan04} for $z\approx 6$ galaxies, which have $\approx 0.3$ galaxies per square arcmin at AB=26 mag in approximately the same waveband.  We have therefore assumed an unabsorbed magnitude of AB=26 mag, or $H=24.6$ Vega mag (using the conversion given by \citealt{tok05}).

\begin{figure*}
\centering \includegraphics[scale=0.7]{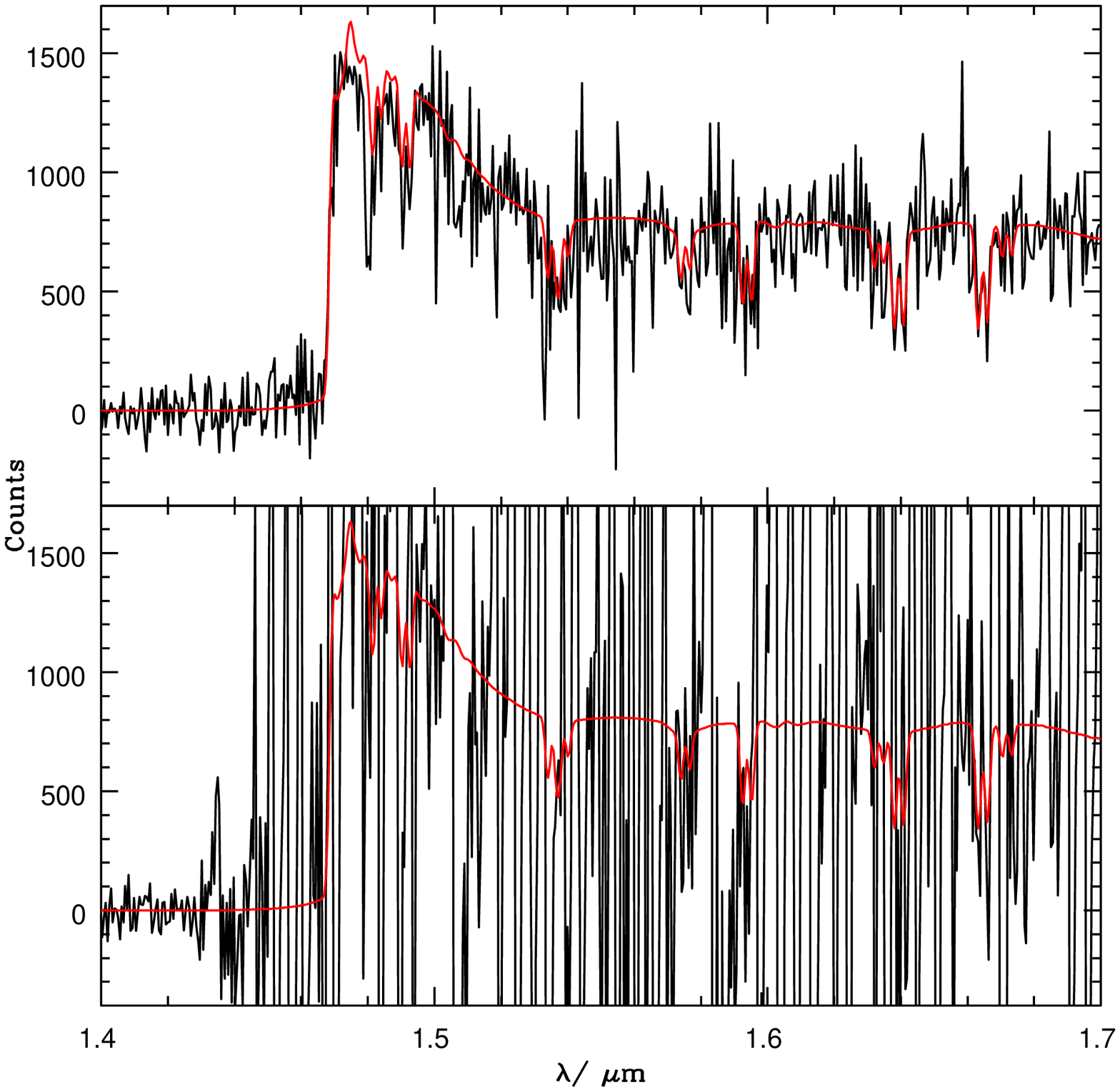}
\caption{A simulated spectrum of an $H=24.6$ Vega mag, $z=11$ QSO as observed by the system described in \S~\ref{sec:sys}.  The exposure time was 70hr, and the spectral resolution was $R=1000$.  The top panel shows a system with FBGs and the bottom panel shows an identical system without FBGs.  The red lines indicate the true object spectrum.}
\label{fig:qsosim}
\end{figure*}

We used a simulated $z=10$, $H=20$ mag spectrum (provided by Richard MacMahon, private communication), which includes \citet{gun65} absorption and metal line absorption.  We then redshifted this to $z=11$ (assuming no further absorption due to neutral hydrogen), and normalised the $H$ band flux to 24.6 mag.  
Figure~\ref{fig:qsosim} shows a simulation of a 70hr exposure at $R=1000$.  The bottom panel shows the results of an identical system without FBGs, and is dominated by residual sky lines.  The use of FBGs results in a remarkable improvement in the quality of the resulting spectrum.  The \citet{gun65} absorption trough is clearly visible when FBGs are used, whereas without FBGs the spectrum is completely dominated by noise and residual sky lines.  FBGs offer the chance to accurately measure the redshifts of extremely faint objects during the epoch of reoinisation.  This will be discussed further in the summary.

\subsubsection{Redshift 3 galaxy simulation}

\begin{figure*}
\centering \includegraphics[scale=0.7]{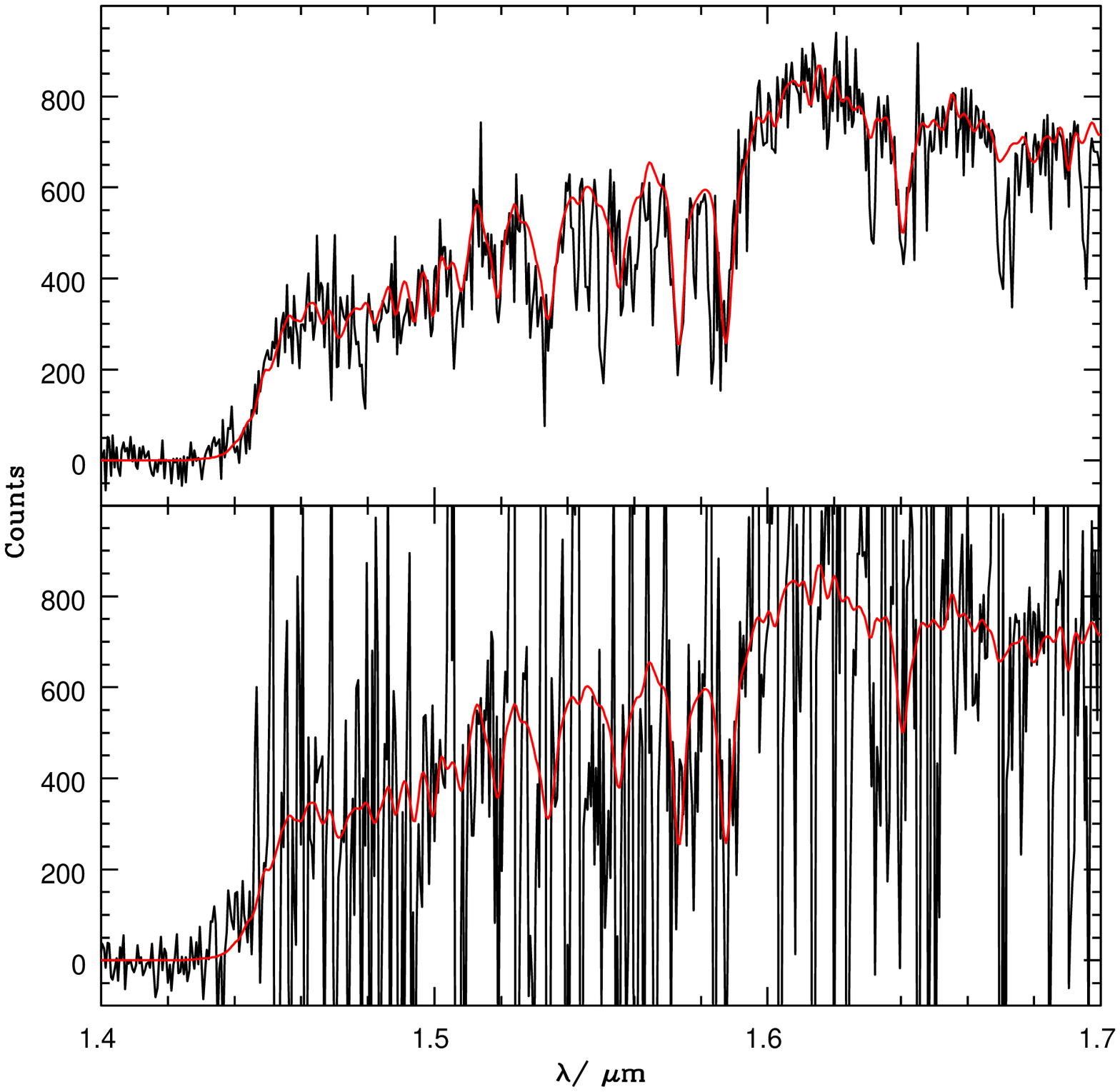}
\caption{A simulated spectrum of an $H=23$ Vega mag, $z=3$ galaxy as observed by the system described in \S~\ref{sec:sys}.  The exposure time was 8hr, and the spectral resolution was $R=1000$.  The top panel shows a system with FBGs and the bottom panel shows an identical system without FBGs.  The red lines indicate the true object spectrum.}
\label{fig:galsim}
\end{figure*}

We have simulated an observation of an $H=23$ Vega mag galaxy at $z=3$.  The exposure time was 8hrs and the resolution was $R=1000$.  The galaxy spectrum is from the synthetic spectral libraries of \citet{bru03}, and assumes an instantaneous burst of star-formation, an age of 1Gyr and solar metallicity.  Figure~\ref{fig:galsim} shows the resulting spectrum as observed through a system with FBGs and a system without FBGs.  The FBGs result in a high quality spectrum with Ca H and Ca K absorption lines clearly visible.  Thus FBGs offer the chance to obtain precise measurements of galaxies when the Universe was only a fifth of its present age, allowing a detailed study of evolution over almost the entire history of galaxies. 

\subsubsection{T dwarf simulation}

Cool stars emit most of their radiation in the NIR and are identified by molecular absorption features.  The faintest T dwarf to date has a magnitude of $J=20.28$ Vega mag (\citealt{zap02}; see Adam Burgasser's T dwarf page\footnote{http://web.mit.edu/ajb/www/tdwarf/}), however most known T dwarfs have magnitudes around $J\sim15$--16 Vega mag, and are located within a few tens of parsecs from the Sun.

\begin{figure*}
\centering \includegraphics[scale=0.7]{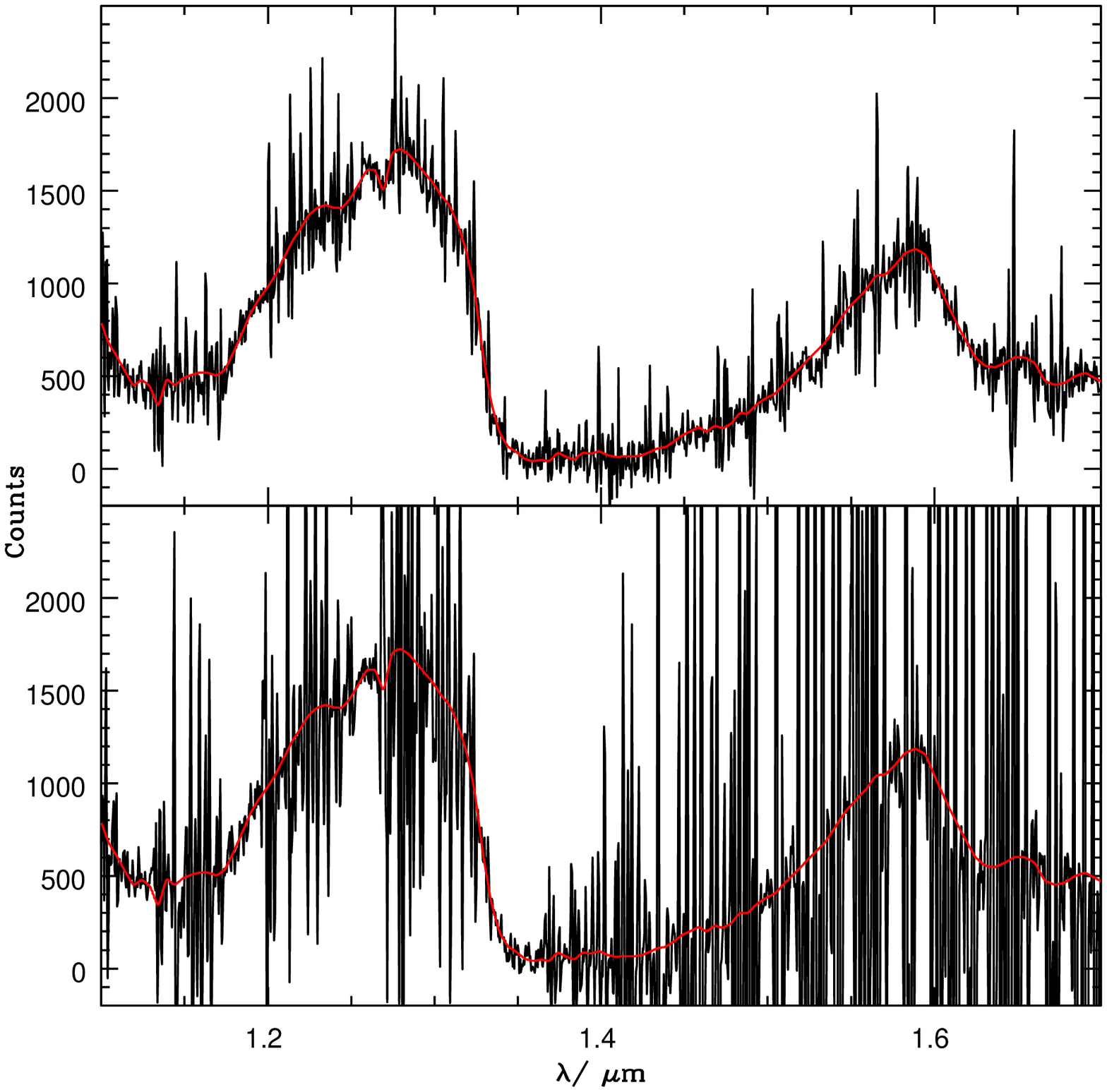}
\caption{A simulated spectrum of a $J=21$ Vega mag,T5 dwarf as observed by the system described in \S~\ref{sec:sys}.  The exposure time was 8hr, and the spectral resolution was $R=1000$.  The top panel shows a system with FBGs and the bottom panel shows an identical system without FBGs.  The red lines indicate the true object spectrum.}
\label{fig:t5sim}
\end{figure*}

Figure~\ref{fig:t5sim} shows a simulated spectrum of a $J=21$ Vega mag T dwarf observed at $R=1000$, in  J and H simulataneously, for 8hrs.  The spectrum is that of a T5 dwarf, 2MASS0559-1404 (\citealt{mcl03}\footnote{http://www.astro.ucla.edu/$\sim$mclean/BDSSarchive/}) scaled to $J=21$.  The FBG spectrum is of much superior quality to the spectrum employing ordinary fibres.   FBGs offer the opportunity to identify much fainter cool stars than ever before.
 
\subsection{Summary}
\label{sec:summary}

We have demonstrated the possibility to achieve extremely deep near-infrared observations using fibre Bragg gratings.  Hitherto, attempts to obtain deep NIR observations have been frustrated by the extremely bright and variable OH emission from the atmosphere.  We have shown that in order to efficiently remove OH lines from astronomical spectra the OH lines must be suppressed prior to the light entering the camera.  Scattering from dispersive optics results in Lorentzian line profiles, the wings of which are much brighter than the true interline continuum, and in most cases are brighter than the sources being observed.  The variability of the OH lines makes it extremely difficult to subtract cleanly.

FBGs can suppress OH lines at high resolution, $R=10,000$, high suppression (30dB) and with high throughput out of the suppression regions ($\approx 90$ per cent).  Suppressing 150 lines in the $J$ or $H$ band, with a maximum suppression of 30dB, decreases the sky background by a factor of $\approx 45$, or 4 magnitudes.  Achieving this will have the potential to extend NIR observation to objects orders of magnitude fainter than currently possible. 
The resulting observations will  have a great impact on many areas of astronomy, and we highlight a few now.

\subsubsection{First light and reionisation}

FBGs offer the opportunity to study the rest-frame UV light from objects at the time of reionisation.  Figure~\ref{fig:qsosim} shows a simulated spectrum of a $z=11$ QSO with a clearly visible \citet{gun65} absorption trough.  Spectroscopy with FBGs would therefore be ideally suited to identifying first light sources, performing follow-up spectroscopy from surveys of \lya\ emission (e.g.\ \citealt{tan05}), continuum drop-out due to the Lyman-break (e.g.\ \citealt{sta04}) and ultra-narrow band surveys (e.g.\ \citealt{hor04}).  A spectrum of a complete Gunn-Peterson trough would confirm the redshift of the source, which would allow estimates of the luminosity function and number-magnitude relations of the first sources of light in the Universe.  If the sources are bright enough, then it may be possible to also measure  the line profile of \lya\, which is expected to be distinctively asymmetrical due to resonant scattering bluewards of the line-centre.  The intrinsic width of the line itself will depend upon the type of source; quasars will have broad \lya\ lines, FWHM$\approx 1500$km s$^{-1}$, whereas galaxies will have narrower lines, FWHM$\approx 100$km s$^{-1}$.  The difference in line widths leads to the possibility of distinguishing the type of sources responsible for ionisination.  For the purpose of confirming the presence of a redshifted \lya\ line it is necessary to resolve the lines sufficiently to measure the asymmetry, requiring resolutions of $R\gsim 300000/100 =  3000$.  

The combination of FBGs and  ELTs will make it possible to thoroughly explore the ionisation history of the IGM using NIR spectroscopy of first light objects.  The detection of complete Gunn-Peterson absorption troughs in the spectra of high redshift quasars (\citealt{bec01}) show that the IGM was partly neutral at $z\approx 6.5$.  However, a neutral fraction of only $x_{\rm HI}\approx 10^{-4}$ is all it takes for complete absorption due to the high oscillator strength of the \lya\ transition.  Thus to probe further back in time to uncover the details of the ionisation of the IGM, we must look to other means.

\citet{oh02} suggest that metal absorption lines may provide a useful means of measuring the ionisation state of the IGM.  O{\sc i} and Si{\sc ii} have ionisation potentials close to that of neutral hydrogen, and thus measuring the neutral fraction of either of these metals could be used to infer the neutral fraction of hydrogen.  Furthermore there are useful lines just longer than \lya\ so a single measured spectrum may incorporate \lya\ 1215.67\AA, O{\sc i} 1302.17\AA\ and Si{\sc ii} 1260.42\AA.  These metals have previously been observed in the spectra of high redshift quasars (\citealt{pet01}) verifying their usability for this purpose.  Since it is unlikely that metals will be smoothly distributed through the IGM, the signature will be in the form of a forest of lines rather than a Gunn-Peterson trough.  A measurement of the equivalent width of the lines, or even simply the number of lines will provide precious information on the ionisation state.  

The detection or non-detection of a radiation damping wing has the potential to divulge important information on the ionisation state of the IGM.  If a radiation damping wing is found it will provide a very accurate measure of the neutral fraction of hydrogen (\citealt{mir98}).  If the ionising sources reside in significant cosmic Str\"{o}mgren spheres then a damping wing will be impossible to measure (\citealt{cen00}; \citealt{mad00}).  However, the line profile may still be identified as \lya\ due to its asymmetrical shape, and thus the lack of a damping wing can place constraints on the ionising fluxes of the sources.  If the sources are galaxies then it may also place constraints on the clustering of the objects, since it is unlikely that a single primordial galaxy would produce a significant ionised region on its own.  If spectra can be obtained with a sufficiently high signal-to-noise, then the tail of the damping wing may provide detailed information on the size of the Str\"{o}mgren spheres.

\begin{figure*}
\centering \includegraphics[scale=0.7]{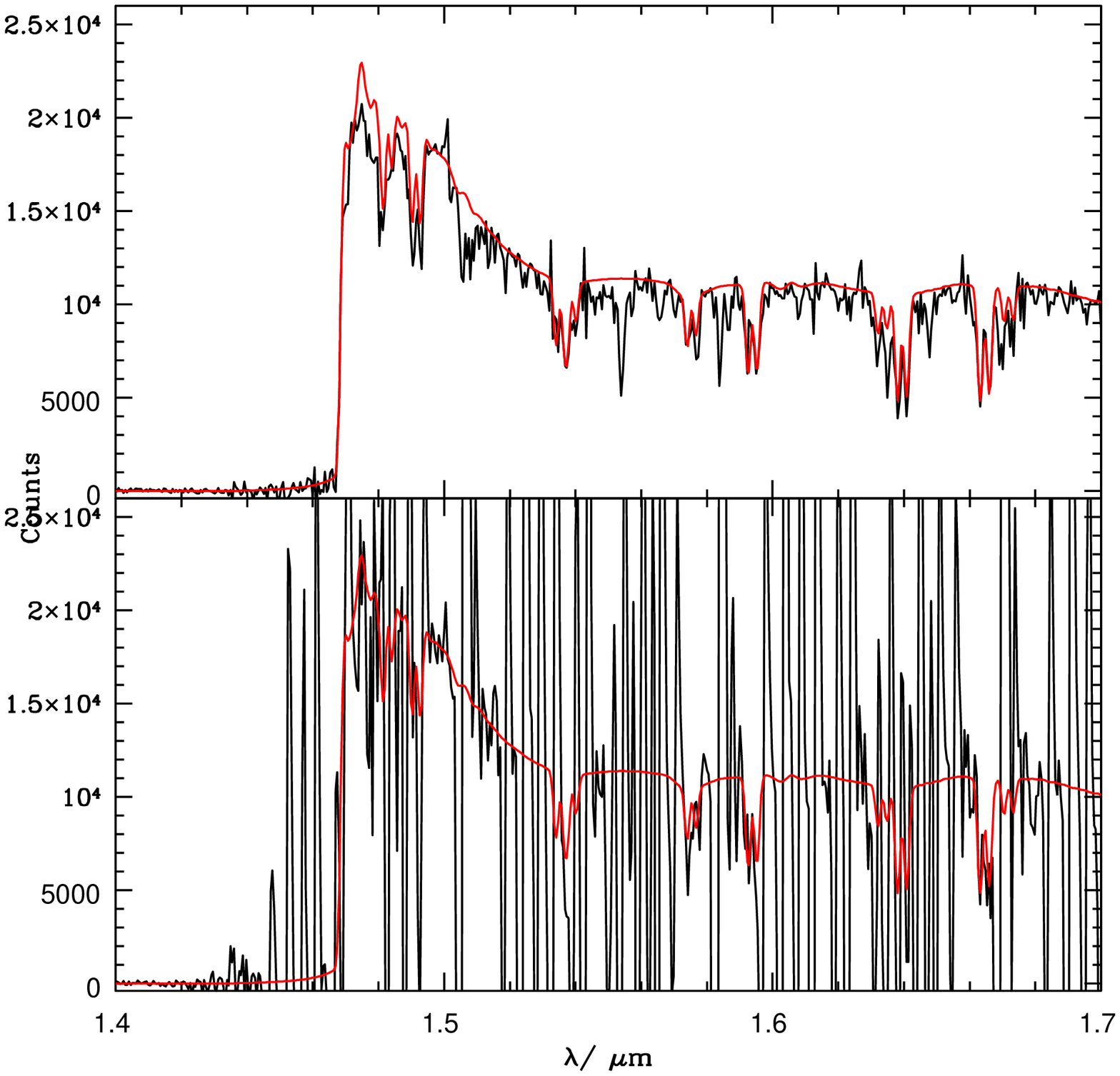}
\caption{A simulated spectrum of an $H=24.6$ Vega mag, $z=11$ QSO as observed by the system described in \S~\ref{sec:sys} with a 30m telesscope.  The exposure time was 70hr, and the spectral resolution was $R=1000$.  The top panel shows a system with FBGs and the bottom panel shows an identical system without FBGs.  The red lines indicate the true object spectrum.}
\label{fig:qso30m}
\end{figure*}

Figure~\ref{fig:qso30m} shows the same simulation as described in section~\ref{sec:qsosim}, except the telescope was assumed to have a diameter of 30m.  Metal absorption lines are now visible in the spectrum.

\subsubsection{Galaxy evolution}

Deep NIR imaging and spectroscopy with FBGs will allow galaxy evolution to be studied over most of the history of the Universe.  Figure~\ref{fig:galsim} shows a $H=23$ Vega mag galaxy at $z=3$.  Absorption lines are clearly visible.  It would thus be possible to measure galaxy masses, luminosities and star-formation rates from emission lines, and to derive the evolution of the star-formation history, the luminosity function, the \citet{fab76} relation and if high spatial resolution imaging is available, the fundamental plane.

\subsubsection{Cool stars}

NIR infrared observations are essential for identifying cool stars such as L, T and Y dwarfs.  Measuring the number density of such objects as a function of mass is a critical test of our understanding of both star and planet formation processes.  NIR spectroscopy is essential for fully characterising their temperature, surface gravity and atmospheric conditions.  The use of FBGs would allow observations of objects orders of magnitude fainter than currently possible (see Figure~\ref{fig:t5sim}), thus extending the survey volumes and numbers of such objects, and placing strong constraints on the low mass end of the initial mass function.

\section{Realisation}
\label{sec:real}

The 8-10m telescope generation has been with us for more than a decade, and astronomy must now plan for a new generation of facilities if the field is to advance in the years to come. For ground-based observations, the EarthÕs atmosphere provides a major challenge to scientific progress on two fronts: (a) the movement of air cells blurs the cosmic signal; (b) hydroxyl emission from the upper atmosphere produces a perpetual twilight, thereby greatly increasing the background signal in astronomical observations. At near infrared wavelengths, there are clear indications that adaptive optics is overcoming the problem of ÒseeingÓ and is starting to deliver unprecedented resolution and sensitivity at near-infrared wavelengths (\citealt{liu06}). In contrast, the extreme brightness of the infrared sky remains an unsolved problem.

The James Webb Space Telescope (\citealt{gar06}) will be launched sometime in the next decade, bypassing altogether the problems associated with observing through the atmosphere.  However, it will have a 6m aperture compared to apertures of 20--42m of the next generation of ground-based telescopes.  The satellite will be inaccessible and will not be
serviced by astronauts, thus the instrument suite will have restricted functionality compared to what is possible today on ground-based sites.  For these reasons it is important to continue to address the issue of the infrared night sky background for ground-based telescopes.
 
The AAO has spearheaded the development of a new technology that promises to completely solve the Òinfrared twilightÓ problem, i.e.\ by removing 99 per cent of the OH line emission with an overall efficiency of 80 per cent or better. The concept exploits two recent advances in photonics: (i) broadband fibre Bragg gratings (FBGs); (ii)  multimode to single-mode photonic lanterns. While the development remains challenging, there are indications that the development is on track to solve the OH suppression problem once and for all. We have demonstrated the extraordinary scientific potential of these fibres in this paper.  Note that all simulations include correction by adaptive optics, for both cases, with and without FBGs.  The observations we have presented and discussed would not be possible through use of adaptive optics alone.  The improvements over the AO-only observations demonstrate a compelling need for FBGs, if such deep observations are to be achieved.  The development and implementation of FBG fed instruments on existing 8m, and especially on future ELTs, deserves equal  attention to that which has been invested in the development of adaptive optics.

In subsequent papers, we will discuss the relative merits of OH suppression in natural seeing compared
to corrected seeing through adaptive optics, and describe instrument concepts that are under
development to capitalize on the photonic OH suppression development.

\section*{Acknowledgments}

We thank Adam Burgasser for providing the T dwarf spectrum, and Richard McMahon for the QSO spectrum, which were used in the simulations.  We thank members of the instrument science group at the AAO for invaluable discussions.  We thank the referee for useful comments which have improved this paper.
This research was funded by the STFC grant PP/D002494/1 Revolutionary Technologies for Optical-IR Astronomy.  JBH is supported by a Federation Fellowship from the Australian Research Council.
 
\bibliographystyle{scemnras}
\bibliography{clusters}

\end{document}